%% file: 3g_2.tex
\documentclass[preprint,aps,prd,letterpaper,tightenlines,nofootinbib,showpacs]{revtex4-1}
\usepackage{amsmath,amssymb,graphicx,mathrsfs}
\usepackage[usenames,dvipsnames]{xcolor}
\usepackage{dcolumn}
\usepackage{bm}
\usepackage{feynmp}
\usepackage{ifpdf}
\ifpdf
\fi
\makeatletter
\def\endfmffile{%
  \fmfcmd{\p@rcent\space the end.^^J%
          end.^^J%
          endinput;}%
  \if@fmfio
    \immediate\closeout\@outfmf
  \fi
  \IfFileExists{\thefmffile.mp}{\immediate\write18{mpost \thefmffile}}{}
  \let\thefmffile\relax
}
\makeatother
\newcommand{\q}{\bar{q}}
\newcommand{\x}{{\bf x}}
\newcommand{\y}{{\bf y}}

\newcommand{\pslash}{\not\hspace{-0.7mm}p}

\newcommand{\be}{\begin{equation}}
\newcommand{\ee}{\end{equation}}
\newcommand{\nn}{\nonumber\\}
\newcommand{\eqn}[1]{\label{#1}}
\newcommand{\eq}[1]{Eq.~(\ref{#1})}
\newcommand{\eqs}[1]{Eqs.~(\ref{#1})}
\newcommand{\fign}[1]{\label{#1}}
\newcommand{\fig}[1]{Fig.~\ref{#1}}

\newcommand{\bPhi}{\bar{\Phi}}

\newcommand{\la}{\langle}
\newcommand{\ra}{\rangle}

\begin{document}
\title{Gauge invariant formulation of \bm{$3\gamma$} decay of particle-antiparticle bound states}
\author{B. Blankleider}
\affiliation{
School of Chemical and Physical Sciences,
 Flinders University, Bedford Park, SA 5042, Australia}
\email{boris.blankleider@flinders.edu.au}
\author{A. N. Kvinikhidze}
\affiliation{A.\ Razmadze Mathematical 
Institute, Georgian Academy of Sciences, Merab Aleksidze St.\ 1, 380093 Tbilisi, Georgia}
\email{sasha\_kvinikhidze@hotmail.com}
\author{Z. K. Silagadze}
\affiliation{
Budker Institute of Nuclear Physics, SB RAS, Novosibirsk 630090, Russia}
\affiliation{Novosibirsk State University, Novosibirsk 630090, Russia}
\email{silagadze@inp.nsk.su}



\begin{abstract} 
We construct the gauge invariant three-photon decay amplitude of particle-antiparticle bound states modeled by the Dyson-Schwinger and  Bethe-Salpeter equations. Application to the quark-antiquark ($q\q$) bound states is emphasized.  An essential aspect of our formulation is that it applies to any underlying quantum field theoretic model of the $q\q$ system, and not just to models, like exact QCD, where the quark self-energy $\Sigma$ couples to the electromagnetic field solely via dressed quark propagators. In this way, applications to effective field theories and other QCD motivated models are envisioned. The three-photon decay amplitude is constructed by attaching currents to all possible places in the Feynman diagrams contributing to the dressed quark propagator. The gauge invariance of our construction is thus a direct consequence of respecting the underlying structure of the quantum field theory determining the dynamics. In the resultant 
expression for the three-photon decay amplitude, all the basic ingredients consisting of  the bound state wave function, the final-state interaction $q\q$ $t$ matrix, the dressed quark propagator, and dressed quark currents, are determined by a universal Bethe-Salpeter kernel. 
\end{abstract}

\maketitle


\section{Introduction}

In this paper we derive the gauge invariant amplitude for $\Phi\rightarrow 3\gamma$ decay in the context of relativistic quantum field theory (QFT).
Note that $\Phi$ here denotes a quark-antiquark or lepton-antilepton bound state (with appropriate quantum numbers) described by the Bethe-Salpeter (BS) equation \cite{Salpeter:1951sz}  (for a pedagogical
introduction see, for example, \cite{Lurie:1968zz}). A feature of our approach is that for any given model of the BS kernel, gauge invariance is implemented in the way prescribed by QFT (with photons attached to all possible places in the necessarily nonperturbative model).

A well known example of the $\Phi\rightarrow 3\gamma$ transition is supplied by orthopositronium whose three-photon decay provides a laboratory for precision quantum electrodynamics (QED) tests \cite{Karshenboim:2003vs,Asai:2008em}. Analogous decays of $^3S_1$ quarkonia are much more difficult to study both experimentally and  theoretically. On the experimental side, the problem is tiny decay  rates. Unlike orthopositronium, which almost exclusively decays into the three-photon final state, quarkonia have many decay channels and the  three-photon decay rate, being proportional to $\alpha^3$, is very small.  Only recently has the $J/\psi\to 3\gamma$ decay been observed by the CLEO collaboration \cite{cleo}, and the accuracy of the branching fraction determination, further improved in the BES-III experiment \cite{bes-iii}. Despite tiny decay rates, three-photon decays of vector quarkonia are of  considerable theoretical interest as they can provide valuable information about the $q\q$ bound state dynamics in the context of quantum chromodynamics (QCD) \cite{Voloshin:2007dx}.

On the theoretical side, considerable success of nonrelativisitic QED \cite{Caswell:1985ui} in positronium studies stimulated similar investigations of quarkonia properties in the framework of nonrelativistic QCD \cite{Caswell:1985ui,Brambilla}. However, in the case of $J/\psi\to 3\gamma$ decay, there has been a struggle to reconcile  theory with experiment \cite{Feng:2012by}. The lowest order prediction is several times greater than the experimental result and the inclusion of the known ${\cal O} (\alpha_s)$ and ${\cal O}(v^2)$ corrections only worsens the situation, even making the theoretical prediction of dubious  physical significance \cite{Feng:2012by}. For light quarkonia like $\omega$ and $\phi$ mesons, we expect even more severe theoretical challenges as the  nonrelativistic QCD completely breaks down and fully fledged relativistic bound state theory is needed for their description.

There is also the following interesting theoretical problem in  $\Phi\rightarrow 3\gamma$ decay related to gauge invariance \cite{Smith}.  As is well known, gauge invariance constrains the structure of amplitudes  as it implies the fulfilment of Ward-Takahashi (WT) identities \cite{Ward:1950xp,Takahashi:1957xn, Bentz:1986nq}. F.~E.~Low realised  long ago \cite{Low:1958sn} that these identities, combined with the analyticity of  the amplitudes, restricts severely the first two terms of the Laurent  series expansion of the amplitude in the external photon energy. The resulting Low's theorem is quite a general model-independent result, and implies that the $\Phi\rightarrow 3\gamma$ decay amplitude must vanish linearly  in soft photon energy when the energy of one of the photons is going to zero. Interestingly, Low's theorem is violated for all currently popular orthopositronium decay amplitudes \cite{Smith,Smith1}.

The root of the problem is well understood \cite{Smith,Smith1}. Current approaches connect the $\Phi\rightarrow 3\gamma$ decay amplitude to the amplitude of annihilation of free asymptotic charged particles, as the charged 
particles in the bound state are considered quasi-free. But this procedure introduces spurious infrared singularities via bremsstrahlung-type radiation, absent for true bound states as the neutral boson cannot radiate very soft photons which do not resolve the internal structure of the boson. While the  infrared singularity itself is canceled in the decay rate, it leaves a ${\cal O}(1)$  imprint in the decay rate which spoils the consistency with  Low's theorem.
For orthopositronium, the binding energy effects are very small and this  violation of Low's theorem has probably only academic interest at the present  level of experimental precision. However for quarkonia, binding energy 
effects are important, so it is desirable to have a general, theoretically correct procedure for constructing gauge invariant amplitudes that are compatible with the Bethe-Salpeter formalism for bound states. 

For this purpose, we use 
 the "gauging of equations method" \cite{Haberzettl:1997jg,KB1,KB2,KB3,KB4,light-fr} to express the $\Phi\rightarrow 3\gamma$  amplitude
$\la0|TJ^\sigma(y)J^\nu(x)J^\mu(0)|\Phi\ra$ in terms of the $\Phi$-meson quark-antiquark 
($q\bar q$) bound state wave function,\footnote{The results of this paper apply to any particle-antiparticle bound state that can decay to three photons; however, for the purposes of presentation, we use the example of the $q\q$ system whose bound state is a generic "$\Phi$-meson". }  the $q\bar q$ scattering amplitude, the quark propagator,  the quark-photon vertex, and  the gauged kernels $K^\mu$, $K^{\mu\nu}$, and $K^{\mu\nu\sigma}$.  All these input quantities are determined by the same model BS kernel $K$.
Gauge invariance in this approach results from respecting the general structure of QFT in that the $(n+1)$-point Green function ${\cal G}^\mu$, which involves a current operator $J^\mu$ in addition to all fields involved in the corresponding $n$-point Green function ${\cal G}$, is constructed
by attaching a current line (labeled by $\mu$) to all possible places in every Feynman diagrams contributing to ${\cal G}$.
Such consistency can be seen already at the first step of our 
construction in Section \ref{phi-1g},  where we derive the amplitude $A^\mu=\la0|J^\mu(0)|\Phi\ra$ for transition of a meson to a virtual photon, $\Phi\rightarrow \gamma$. In this case, formally removing the current operator $J^\mu(0)$ from the corresponding Feynman diagrams of $\la0|J^\mu(0)|\Phi\ra$, results in a matrix element $\la0|\Phi\ra$ which is zero. We therefore attach photons instead to the dressed quark propagator $S$ by
gauging the Dyson-Schwinger (DS) equation for $S$. In this way we obtain the 3-point Green function $S^\mu$
corresponding to quark-antiquark annihilation into a virtual photon, $q\bar q\rightarrow \gamma$.
Then the $\Phi\rightarrow \gamma$ amplitude, $A^\mu$, is extracted by taking the residue at the pole of $S^\mu$ corresponding to a quark-antiquark bound state located at $P^2=M^2_\Phi$.
In Section \ref{phi-2g} we gauge the amplitude $A^\mu$   to obtain the amplitude $A^{\mu\nu}$ for $\Phi\rightarrow 2\gamma$, and similarly in Section \ref{phi-3g} we gauge the amplitude $A^{\mu\nu}$   to obtain the amplitude $A^{\mu\nu\sigma}$ for $\Phi\rightarrow 3\gamma$.
The derived expressions for $A^{\mu\nu}$ and $A^{\mu\nu\sigma}$  are gauge invariant by construction, as they are obtained by attaching photons everywhere in the amplitudes $A^{\mu}$ and $A^{\mu\nu}$ , respectively.

The successfully developing continuum QCD approach \cite{Alkofer:2000wg,  Maris:2003vk, Fischer:2006ub, Roberts:2007jh,  Eichmann:2009zx, Eichmann:2011ec, SanchisAlepuz:2012vb}, which is based on the DS equations of QCD, is a natural framework for further application of the results presented in this paper. 

\section{Formulation}

The amplitude for the one-photon disintegration process $\Phi\rightarrow \gamma$ is defined in QFT by
\be
A^{\mu} = \la0|J^\mu(0)|\Phi\ra  \eqn{Amudef}
\ee
where $J^\mu(z)$ is the electromagnetic (EM) current operator. In the case of exact QCD,  $J^\mu(z)$  is given in terms of the quark field $q(z)$ and quark electric charge operator $e$, as
\be
J^\mu(z) =\q(z) e\gamma^\mu q(z) ,
\eqn{Jmu}
\ee
which after substitution into \eq{Amudef}, gives
 \be
A^\mu =\int \frac{d^4p}{(2\pi)^4} \mbox{Tr}\left[e\gamma^\mu\Phi(P,p)\right] \eqn{Amu1}
 \ee
where $\Phi(P,p)$ is the $q\q$ bound state wave function defined by
\be
\Phi(P,p)=\int d^4y\,e^{ip\cdot y}\la 0|Tq(y)\bar q(0)|\Phi\ra. 
\ee
Although this derivation of \eq{Amu1} is strictly valid for the case of exact QCD, one can ask if this expression can also be used for {\em models} of QCD. In particular, we are interested in the validity of \eq{Amu1} for descriptions where
 the bound state wave function is modeled by the BS equation
 \be
 \Phi(P,p)=G_0(P,p)\int \frac{d^4k}{(2\pi)^4}K(P,p,k)\Phi(P,k)   \eqn{BS}
\ee
where $G_0$ is the free $q\q$ Green function and $K$ is a  model BS kernel.  To help answer this question, we appeal to the property of gauge invariance.

Although the expression for $A^\mu$ given by \eq{Amu1} is gauge invariant for the case where the wave function $\Phi$  is the solution of the BS equation using the exact QCD kernel $K$, gauge invariance is generally lost when approximations are made for $K$. More precisely, \eq{Amu1} is gauge invariant only if $K$ satisfies a consistency condition  relating it to the quark self-energy $\Sigma$ (which is used to calculate the $G_0$ in the BS equation). The details of this consistency condition are presented in Appendix A. 

Although it is possible to construct models for $K$ and $\Sigma$ where this consistency condition is satisfied, these are limited to models where the quark self-energy $\Sigma$ is a functional only of the fully dressed quark propagator $S$, as in exact QCD ($\Sigma$ can of course be also a functional of quantities that do not carry charge, like the dressed gluon propagator,  but these contribute only to the transversal component of the current and are therefore not of direct relevance to the question of gauge invariance).
The conclusion, therefore, is that \eq{Amu1} is gauge invariant only for those QCD models whose $\Sigma$ is a functional of just $S$.  It is not gauge invariant for the widely used Nambu-Jona-Lasinio (NJL) model of QCD \cite{Klevansky:1992qe} because there are, for example, four-quark vertices which support electromagnetic charge flow that entail four-quark current vertices on the Lagrangian level.

Here we shall pursue a more general approach that applies to {\em any} model of $\Sigma$, as demonstrated in Appendix A. In this more general approach, the gauging of equations method is used to develop expressions for the gauge invariant photo-decay amplitudes.

\subsection{Gauge invariant  \bm{$\Phi\rightarrow \gamma$} amplitude}\label{phi-1g}

We start with the dressed quark propagator $S(p)$ defined by
\be
(2\pi)^4\delta^4(p'-p) S(p) = \int d^4y\, d^4x\, e^{i(p'y-px)}\la 0 |T q(y) \q(x)|0\ra . \eqn{S}
\ee
One can express $S(p)$ in terms of the bare quark propagator $S_0(p)=i/(\pslash-m_0)$ and the quark self-energy $\Sigma(p)$ as
\be
-S^{-1}(p)= -S_0^{-1}(p) + \Sigma(p),
\ee
which in exact QCD becomes the Dyson-Schwinger (DS) equation
 \be
-[ S^{-1}]^a_b(p)= -[S_0^{-1}]^a_b(p)+\int\frac{d^4k}{(2\pi)^4}V_{bc}^{ad}(p,p;k,k) S^c_d(k) , \eqn{SDexact}
\ee
where $V$ is the asymmetrical one-gluon exchange $q\q$ amplitude with one bare and one dressed quark-gluon vertex, as illustrated in \fig{D}.  
Note that we use a convention where momentum variables in $V_{bc}^{ad}(p_1,p_2;k_1,k_2)$ are assigned according to the direction of the quark lines; i.e., a quark line with assigned momentum $p$ corresponds to a quark moving in the direction of the arrow with momentum $p$, or an antiquark moving in the opposite direction with momentum $-p$. In particular,
$k_1$ and $p_1$ are the momenta of the incoming and outgoing quarks, respectively, while $-k_2$ and $-p_2$ are the momenta of the incoming and outgoing antiquarks. The total momentum of the quark-antiquark system is thus
$P= k_1-k_2=p_1-p_2$. Each of the labels $a, b, c, d$ denotes the set of indices labeling the particles (in the case of quarks it includes the Dirac spinor index). The upper indices, $a$ and $d$, are associated with incoming quarks in the final and initial state, respectively, while the lower indices, $b$ and $c$, are similarly associated with the outgoing quarks. In \eq{SDexact} and all expressions below, identical upper and lower indices are assumed to be summed.
\begin{figure}[t]
\begin{fmffile}{D}
\[
\parbox{40mm}{
\begin{fmfgraph*}(40,30)
\fmfstraight
\fmftopn{f}{3}\fmfbottomn{i}{3}
\fmf{fermion,tension=1.3,l.s=right,label=$p,,a$}{i1,c}
\fmf{fermion,tension=1.3,l.s=right,label=$p',,b$}{c,f1}
\fmf{fermion,tension=1.3,l.s=right,label=$k',,d$}{f3,c}
\fmf{fermion,tension=1.3,l.s=right,label=$k,,c$}{c,i3}
\fmfv{d.s=square,d.f=full,d.si=8}{c}
\end{fmfgraph*}} 
\quad \equiv\quad\quad
\parbox{40mm}{
\begin{fmfgraph*}(40,30)
\fmfstraight
\fmftopn{f}{3}\fmfbottomn{i}{3}
\fmf{fermion,tension=1.3,l.s=right,label=$p,,a$}{i1,ic}
\fmf{fermion,tension=1.3,l.s=right,label=$k,,c$}{ic,i3}
\fmf{fermion,tension=1.3,l.s=right,label=$p',,b$}{fc,f1}
\fmf{fermion,tension=1.3,l.s=right,label=$k',,d$}{f3,fc}
\fmf{gluon,tension=1.3}{fc,ic}
\fmfv{d.s=circle,d.f=full,d.si=8}{ic}
\end{fmfgraph*}} 
\]
\end{fmffile} 
\vspace{-1mm}

\caption{\fign{D} Definition of the asymmetrical one-gluon exchange $q\q$ amplitude $V^{ad}_{bc}(p',p;,k',k)$, symbolised by the black square in the left diagram. The defining Feynman diagram on the right shows a dressed gluon (curly line) connecting at a bare quark-gluon vertex at the top, and a dressed quark-gluon vertex (indicated by the black circle) at the bottom. }
\end{figure}
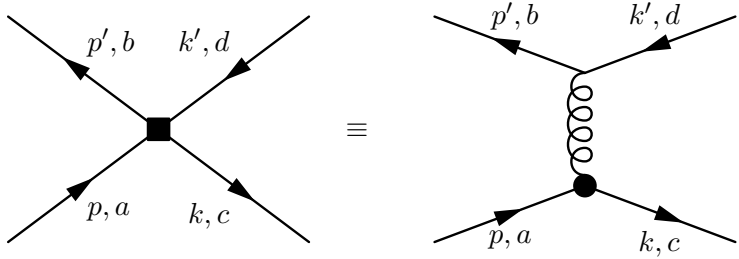

\begin{figure}[b]
\begin{fmffile}{SS}
\[-\ \
\left(\parbox{15mm}{
\begin{fmfgraph*}(15,30)
\fmfstraight
\fmftop{f3,f2,f1}\fmfbottom{i3,i2,i1}
\fmf{fermion,tension=1.3,l.s=left,label=$p$}{i2,v2,f2}
\fmfv{d.s=circle,d.f=full,d.si=7}{v2}
\end{fmfgraph*}} \right)^{-1}_{ba}
=\ \ - \ \
\left(\parbox{15mm}{
\begin{fmfgraph*}(15,30)
\fmfstraight
\fmftop{f3,f2,f1}\fmfbottom{i3,i2,i1}
\fmf{plain_arrow,tension=1.3,l.s=left,label=$p$}{i2,f2}
\end{fmfgraph*}} \right)^{-1}_{ba}
+\quad\quad
\parbox{22mm}{
\begin{fmfgraph*}(22,30)
\fmfstraight
\fmfrightn{r}{3}\fmfleftn{l}{3}
\fmf{fermion,tension=1,l.s=left,label=$p,,b$}{v1,l3}
\fmf{fermion,tension=1,l.s=left,label=$p,,a$}{l1,v1}
\fmf{phantom,tension=25}{v1,v2,v3}
\fmf{phantom,tension=1}{v3,r2}
\fmffreeze
\fmf{fermion,tension=.8,right=1,label=$k,,d$}{r2,v3}
\fmf{fermion,tension=.8,right=1,label=$k,,c$}{v3,r2}
\fmfv{d.s=circle,d.f=full,d.si=7}{r2}
\fmfv{d.s=square,d.f=full,d.si=8}{v2}
\end{fmfgraph*}}
\]
\end{fmffile} 
\vspace{-1mm}

\caption{\fign{fig:DS}Illustration of the Dyson-Schwinger equation, \eq{DS}. Lines decorated with a black circle correspond to dressed quark propagators  while the undecorated line denotes a bare quark propagator. The
black square represents the $q\q$ amplitude $K$.}
\end{figure}
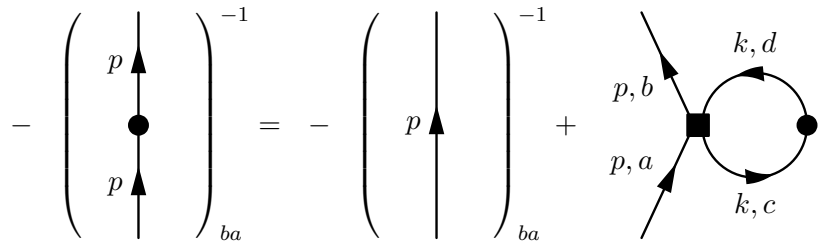

Although the DS equation cannot yet be solved for the case of exact QCD, it provides a convenient starting point for constructing models of $S(p)$. Indeed, we shall assume that the exact amplitude $V$ of \fig{D} is replaced by a model amplitude $K$ for which the DS equation
 \be
-[ S^{-1}]^a_b(p)= -[S_0^{-1}]^a_b(p)+\int\frac{d^4k}{(2\pi)^4}K_{bc}^{ad}(p,p;k,k) S^c_d(k)  \eqn{DS}
\ee
can be solved.\footnote{Although $K$ is defined here as the model DS kernel, it will also be used to model the BS kernel, as explained below. Hence the use of the symbol $K$ for the BS kernel in \eq{BS}.}  Equation (\ref{DS}) is illustrated in \fig{fig:DS}. Implicit in \eq{DS} is that the quark self-energy is modeled as
\be
\Sigma^a_b(p) = \int\frac{d^4k}{(2\pi)^4}K_{bc}^{ad}(p,p;k,k) S^c_d(k).
\ee

The next step is to construct, within the model for $S(p)$, the EM 3-point function $S^\mu$, which in exact QFT is defined by
\begin{align}
S^\mu(p',p)&=\int d^4y\, d^4x\, e^{i(p' y-p x)}\la0|Tq(y)\bar q(x)J^\mu(0)|0\ra \nn
&=\int d^4z\, d^4x\, e^{-i(q z+p x)}\la0|Tq(0)\bar q(x)J^\mu(z)|0\ra \eqn{Smudef}
\end{align}
where $q=p'-p$ is the incoming photon momentum associated with the current operator $J^\mu$.
To do this, we use the gauging of equations method, which when applied to \eq{DS}, gives
\begin{subequations} \eqn{Gammamu-right}
\be
[\Gamma^\mu]^a_b(p)=[F^\mu]^a_b+\int\frac{d^4k}{(2\pi)^4}K_{bc}^{ad}(p',p;k',k) [S^\mu]^c_d(k),
\eqn{Gammamu}
\ee
where
\be
[F^\mu]^a_b(p)=[e\gamma^\mu]^a_b+\int\frac{d^4k}{(2\pi)^4}[K^\mu]_{bc}^{ad}(p',p;k,k) S^c_d(k).
\eqn{Fmu}
\ee
\end{subequations}
In \eq{Gammamu}, $\Gamma^\mu(p)$ is the dressed quark EM vertex function defined by
\be
S^\mu(p',p) \equiv S^\mu(p) = S(p') \Gamma^\mu(p) S(p). \eqn{Gammamudef}
\ee
Equations (\ref{Gammamu}) and (\ref{Fmu}) are illustrated in \fig{fig:DSmu}. 
\begin{figure}[b]
\begin{fmffile}{Smu}
\begin{align*}
&(a) &
\parbox{25mm}{
\begin{fmfgraph*}(25,30)
\fmfstraight
\fmfleftn{l}{3}\fmfrightn{r}{3}
\fmf{fermion,tension=1,l.s=left,label=$p,,a$}{l1,v}
\fmf{fermion,tension=1,l.s=left,label=$p',,b$}{v,l3}
\fmf{photon,tension=1,l.s=left,label=$\leftarrow q$}{r2,v}
\fmfv{d.s=circle,d.f=full,d.si=7}{v}
\end{fmfgraph*}} 
\ \ &=\quad\quad
\parbox{25mm}{
\begin{fmfgraph*}(25,30)
\fmfstraight
\fmfleftn{l}{3}\fmfrightn{r}{3}
\fmf{fermion,tension=1,l.s=left,label=$p,,a$}{l1,v}
\fmf{fermion,tension=1,l.s=left,label=$p',,b$}{v,l3}
\fmf{photon,tension=1,l.s=left,label=$\leftarrow q$}{r2,v}
\fmfv{d.s=square,d.f=full,d.si=8}{v}
\end{fmfgraph*}} \ \
+\quad\quad
\parbox{35mm}{
\begin{fmfgraph*}(35,30)
\fmfstraight
\fmfrightn{r}{3}\fmfleftn{l}{3}
\fmf{fermion,tension=1,l.s=left,label=$p',,b$}{v1,l3}
\fmf{fermion,tension=1,l.s=left,label=$p,,a$}{l1,v1}
\fmf{phantom,tension=25}{v1,v2,v3}
\fmf{phantom,tension=1}{v3,v4}
\fmf{phantom,tension=1}{v4,r2}
\fmffreeze
\fmf{phantom,right,tag=1}{v3,v4}
\fmf{phantom,left,tag=2}{v3,v4}
\fmfposition
\fmfipath{p[]}
\fmfiset{p1}{vpath1(__v3,__v4)}
\fmfiset{p2}{vpath2(__v3,__v4)}
\fmfiv{d.s=circle,d.f=full,d.si=7}{point length(p1)/4 of p1}
\fmfiv{d.s=circle,d.f=full,d.si=7}{point 3*length(p2)/4 of p2}
\fmf{fermion,tension=.8,right=1,label=$k',,d$}{v4,v3}
\fmf{fermion,tension=.8,right=1,label=$k,,c$}{v3,v4}
\fmfv{d.s=circle,d.f=full,d.si=7}{v4}
\fmf{photon,tension=0,l.s=left,label=$\leftarrow q$}{r2,v4}
\fmfv{d.s=square,d.f=full,d.si=8}{v2}
\end{fmfgraph*}}\\[5mm]
&(b) & \parbox{25mm}{
\begin{fmfgraph*}(25,30)
\fmfstraight
\fmfleftn{l}{3}\fmfrightn{r}{3}
\fmf{fermion,tension=1,l.s=left,label=$p,,a$}{l1,v}
\fmf{fermion,tension=1,l.s=left,label=$p',,b$}{v,l3}
\fmf{photon,tension=1,l.s=left,label=$\leftarrow q$}{r2,v}
\fmfv{d.s=square,d.f=full,d.si=8}{v}
\end{fmfgraph*}} 
\ \ &=\quad\quad
\parbox{25mm}{
\begin{fmfgraph*}(25,30)
\fmfstraight
\fmfleftn{l}{3}\fmfrightn{r}{3}
\fmf{fermion,tension=1,l.s=left,label=$p,,a$}{l1,v}
\fmf{fermion,tension=1,l.s=left,label=$p',,b$}{v,l3}
\fmf{photon,tension=1,l.s=left,label=$\leftarrow q$}{r2,v}
\end{fmfgraph*}} \ \
+\quad\quad
\parbox{35mm}{
\begin{fmfgraph*}(35,30)
\fmfstraight
\fmfrightn{r}{3}\fmfleftn{l}{3}
\fmf{fermion,tension=1,l.s=left,label=$p',,b$}{v1,l3}
\fmf{fermion,tension=1,l.s=left,label=$p,,a$}{l1,v1}
\fmf{phantom,tension=25}{v1,v2,v3}
\fmf{phantom,tension=1}{v3,v4}
\fmf{phantom,tension=1}{v4,r2}
\fmffreeze
\fmf{phantom,tension=1}{v3,v4m}
\fmf{phantom,tension=4}{v4m,v4,v4p}
\fmf{phantom,tension=1}{v4p,r2}
\fmffreeze
\fmf{fermion,tension=.8,right=1,label=$k,,d$,width=1}{v4,v3}
\fmf{fermion,tension=.8,right=1,label=$k,,c$}{v3,v4}
\fmfv{d.s=circle,d.f=full,d.si=7}{v4}
\fmf{photon, right=.3,rubout=4,l.s=right,l.d=3,label=$\hspace{1.5cm}\leftarrow q$}{r2,v1}
\fmfv{d.s=square,d.f=full,d.si=8}{v2}
\end{fmfgraph*}}
\end{align*}
\end{fmffile} 
\vspace{0mm}

\caption{\fign{fig:DSmu} (a) Illustration of the gauged DS equation, \eq{Gammamu}, for the dressed quark EM vertex function, $\Gamma^\mu(p)$ (black circle three-point vertices). (b) Illustration of \eq{Fmu} defining the inhomogeneous term  of the DS equation, $F^\mu(p)$ (black square three-point vertex). The wiggly lines denote EM currents associated with incoming photons of momentum $q$ and polarization index $\mu$.  }
\end{figure}
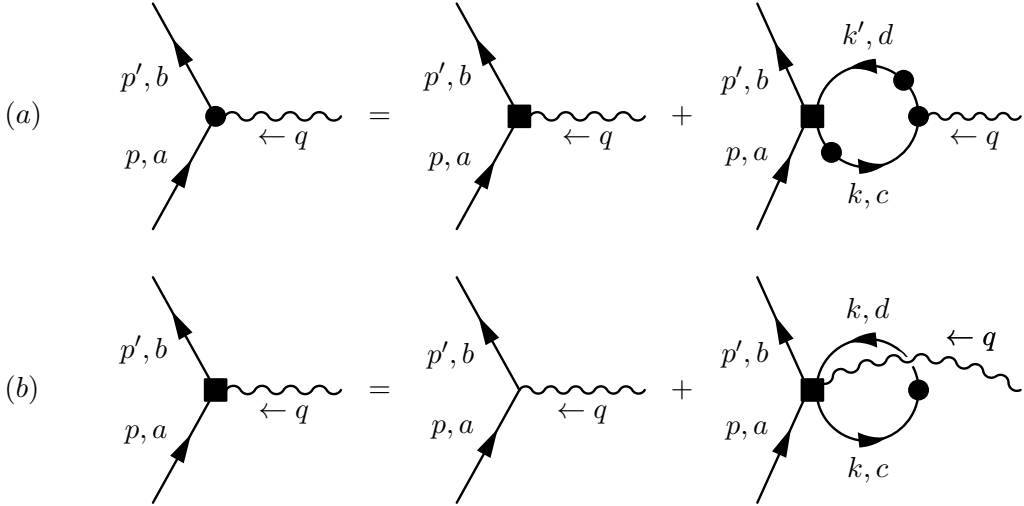
In the above equations we have used an economical notation where the $p'$ dependence of $S^\mu(p)$, $\Gamma^\mu(p)$ and  $F^\mu(p)$ is not written explicitly, but rather, is implied by the photon polarisation index $\mu$; more specifically, we shall associate with index $\mu$ an incoming photon of momentum $q$, so that $p'=p+q$. It is interesting to note that although the DS equation, \eq{DS}, is a nonlinear equation for $S(p)$, the gauged DS equation, \eq{Gammamu}, is a linear equation for $S^\mu(p)$ [or equivalently, for $\Gamma^\mu(p)]$. 

Although $K^\mu$ is a linear functional of $S^\mu$, one cannot yet evaluate its coefficient function in the full theory; however, in most models of interest, $K^\mu$ is either zero, as in the seminal model of Ref.\  \cite{Maris:1999nt}, or does not depend on $S^\mu$ at all (correspondingly $\Sigma$ is a linear functional of $S$), as in the nonlocal versions of the Nambu-Jona-Lasinio (NJL) model  in the leading order (LO) of the $1/N$ expansion  \cite{Ito:1991pv, Plant:1997jr, GomezDumm:2006vz}. In any case, we shall assume that  for the chosen model, the gauged DS equation, \eq{Gammamu}, can be solved for $S^\mu$. Then, writing \eq{Gammamu} in a symbolic notation as\footnote{Such symbolic equations are used throughout this paper. They provide a convenient shorthand where between any two adjacent symbols there are implied  integrals of the form $d^4 k/(2\pi)^4$, taken over independent intermediate momenta (constrained by total momentum conservation), and implied sums over all indices labeling the quarks and antiquarks in the intermediate state.}
\be
\Gamma^\mu = F^\mu + K G_0 \Gamma^\mu, 
\ee
where $G_0(p',p;k',k) = (2\pi)^4 \delta^4(p-k)S(k') S(k)$, it follows that
\be
\Gamma^\mu = (1-KG_0)^{-1} F^\mu.  \eqn{Fmut}
\ee
Denoting by $T$ the $t$ matrix corresponding to the DS kernel $K$, so that
\be
T = K + K G_0 T,   \eqn{BST}
\ee
\eq{Fmut} can be written as
\be
\Gamma^\mu = (1+TG_0) F^\mu  \eqn{Fmuf}
\ee
where
\begin{align}
F^\mu &= e\gamma^\mu + K^\mu S.  \eqn{Fmu2}
\end{align}
The form of $\Gamma^\mu$ expressed by \eqs{Fmuf} and (\ref{Fmu2}) agrees with that given in Ref.\ \cite{Plant:1997jr}.
As \eq{Gammamudef} can be expressed as $S^\mu = G_0 \Gamma^\mu$, one obtains
\be
S^\mu =  G_K F^\mu    \eqn{SmuGF}
\ee
where $G_K$ is the model $q\q$ Green function defined by
\be
G_K = G_0 + G_0 T G_0.   \eqn{BSG}
\ee
In a similar way, one can obtain the conjugate version of \eqs{Gammamu-right}
\begin{subequations} \eqn{Gammamu-left}
\begin{align}
[\Gamma^\mu]^a_b(p)&=[\bar F^\mu]^a_b+\int\frac{d^4k}{(2\pi)^4} [S^\mu]^c_d(k) K_{cb}^{da}(k,k';p,p') ,
\eqn{Gammamu-full}
\end{align}
where
\begin{align}
[\bar F^\mu]^a_b(p)&=[e\gamma^\mu]^a_b+\int\frac{d^4k}{(2\pi)^4}S^c_d(k)[K^\mu]_{cb}^{da}(k,k;p,p'),
\eqn{bFmu-full}
\end{align}
\end{subequations}
or symbolically,
\begin{subequations} \eqn{Gammamu-left-symb}
\begin{align}
\Gamma^\mu &=\bar F^\mu + S^\mu K, \eqn{Gammamy-symb} \\
\bar F^\mu &= e\gamma^\mu + S K^\mu. \eqn{bFmu}
\end{align}
\end{subequations}
Equation (\ref{bFmu-full}) is illustrated in \fig{fig:Fbar}. Similarly, one obtains  the conjugate forms of \eqs{Fmuf} and (\ref{SmuGF}):
\be
\Gamma^\mu =  \bar F^\mu (1+G_0 T)  \eqn{Gammamu-fsi}
\ee
and
\be
S^\mu = \bar F^\mu  G_K .  \eqn{SmuFG}
\ee
\begin{figure}[t]
\begin{fmffile}{Fbar}
\begin{align*}
\parbox{25mm}{
\begin{fmfgraph*}(25,30)
\fmfstraight
\fmfleftn{l}{3}\fmfrightn{r}{3}
\fmf{fermion,tension=1,l.s=left,label=$p,,a$}{r3,v}
\fmf{fermion,tension=1,l.s=left,label=$p',,b$}{v,r1}
\fmf{photon,tension=1,l.s=left,label=$q\rightarrow$}{l2,v}
\fmfv{d.s=square,d.f=full,d.si=8}{v}
\end{fmfgraph*}} 
\ \ =\quad
\parbox{25mm}{
\begin{fmfgraph*}(25,30)
\fmfstraight
\fmfleftn{l}{3}\fmfrightn{r}{3}
\fmf{fermion,tension=1,l.s=left,label=$p,,a$}{r3,v}
\fmf{fermion,tension=1,l.s=left,label=$p',,b$}{v,r1}
\fmf{photon,tension=1,l.s=left,label=$q\rightarrow$}{l2,v}
\end{fmfgraph*}} \ \quad
+\quad
\parbox{35mm}{
\begin{fmfgraph*}(35,30)
\fmfstraight
\fmfrightn{r}{3}\fmfleftn{l}{3}
\fmf{fermion,tension=1,l.s=left,label=$p',,b$}{v1,r1}
\fmf{fermion,tension=1,l.s=left,label=$p,,a$}{r3,v1}
\fmf{phantom,tension=25}{v3,v2,v1}
\fmf{phantom,tension=1}{v4,v3}
\fmf{phantom,tension=1}{l2,v4}
\fmffreeze
\fmffreeze
\fmf{fermion,tension=.8,right=1,label=$k,,c$}{v3,v4}
\fmf{fermion,tension=.8,right=1,label=$k,,d$}{v4,v3}
\fmfv{d.s=circle,d.f=full,d.si=7}{v4}
\fmf{photon, left=.3,rubout=4,l.s=left,l.d=3,label=$\hspace{-1.5cm}q\rightarrow$}{l2,v3}
\fmfv{d.s=square,d.f=full,d.si=8}{v2}
\end{fmfgraph*}}
\end{align*}
\end{fmffile} 
\vspace{0mm}

\caption{\fign{fig:Fbar} Illustration of $\bar F^\mu(p)$ as given by \eq{bFmu-full}. All notation has been defined in the previous diagrams.}
\end{figure}
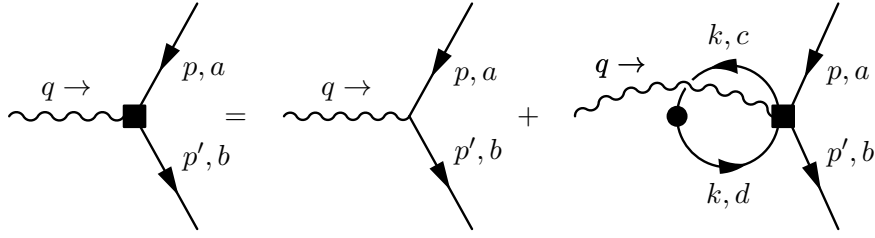
In exact QFT where $S^\mu$ is given by \eq{Smudef}, the  $\Phi\rightarrow \gamma$ amplitude $A^\mu$, defined by \eq{Amudef}, can be extracted from the residue at the pole of $S^\mu(p',p)$ corresponding to a quark-antiquark bound state located at $P^2=M^2_\Phi$:
\begin{align}
S^\mu(p',p) \ & \underset{P^2\rightarrow M_\Phi^2}{\sim} \  i
\frac{\la0|J^\mu(0)|\Phi\ra}{P^2-M_\Phi^2}\bar\Phi(P,p)  \eqn{pole-str}
\end{align}
where $P=p-p'$, and $\bar\Phi(P,p)$ is the conjugate bound state wave function in  momentum space, defined by
\be
\bar\Phi(P,p)=\int d^4y\,e^{-ip\cdot y}\la\Phi|Tq(0)\bar q(y)|0\ra.
\ee
In this respect it is useful to note that the exact QFT $q\q$ Green function $G$, is defined by
\begin{align}
 (2\pi)^4 &\delta^4(k_1-k_2-p_1+p_2) G(k_1,k_2;p_1,p_2)\nn[1mm]
 &=\int d^4y_1\,d^4y_2\,d^4x_1\,d^4x_2\, e^{i(k_1y_1-k_2y_2-p_1x_1+p_2x_2)}
\la0|Tq(y_1)\bar q(y_2)\bar q(x_1)q(x_2)|0\ra,
\end{align}
behaves in the vicinity of the $\Phi$ bound state pole as
\begin{align}
G(k_1,k_2;p_1,p_2) \ & \underset{P^2\rightarrow M_\Phi^2}{\sim} \  i
\frac{\Phi(P,k_1)\bPhi(P,p_1)}{P^2-M_\Phi^2}  \eqn{pole-G}
\end{align}
where $P=p_1-p_2=k_1-k_2$.

To determine $A^\mu$ in the case of models, we use \eq{SmuFG}  to specify $S^\mu$, and then follow the same procedure as for exact QFT; namely, we extract $A^\mu$ from the residue of $S^\mu$ at $P^2 = M_\Phi^2$ in the way implied by \eq{pole-str}. 
In exact QCD such a pole arises from the fact that $K^\mu$ is a functional of $S^\mu$.  By contrast, in  the case of models where $K$ does not depend on the dressed quark propagator $S$, the interaction current $K^\mu$ will correspondingly not depend on $S^\mu$.
Indeed, in this paper we shall assume that the model $K^\mu$ does not have a pole at $M_\Phi$, as this is almost certainly the case in practise. Such a $K$ can be chosen phenomenologically or as a truncated sum of Feynman diagrams; in either case, the only way for the  RHS of \eq{SmuFG}  to have a pole at mass $M_\Phi$, is for the model $K$ to be adjusted so that $M_K=M_\Phi$, which effectively identifies the DS kernel $K$ with the BS kernel and the Green function $G_K$ with $G$. We shall assume such is the case, and therefore drop the subscript $K$ from $G_K$,
writing  \eq{SmuFG}, for example, as\footnote{One should, however, be careful not to confuse \eq{Smu_model}, which applies to {\em models} of QCD, with the equation $S^\mu = e\gamma^\mu G$, which follows from \eq{Smudef} for the case of {\em exact} QCD.}
\begin{align}
S^\mu &= \bar F^\mu G \nn
&= (e\gamma^\mu + S K^\mu) G.  \eqn{Smu_model}
\end{align}
Taking residues of \eq{Smu_model} at $P^2=M_\Phi^2$, we obtain the expression for the model amplitude $A^\mu$:
\begin{align}
A^\mu &= \bar F^\mu \Phi\nn
&=(e\gamma^\mu + S K^\mu) \Phi      \eqn{Am-symbol}
\end{align}
where $K$ is linked to $\Phi$ through the bound state equation (\ref{BS}). 

Equation (\ref{Am-symbol}) has been derived by gauging the DS equation in order to guarantee gauge invariance,  yet at first sight,  it looks inconsistent with \eq{Amu1} due to the extra term $S K^\mu$. This difference, however, is just a reflection of the essential difference between exact QCD whose $K^\mu$ has a pole at $M_\Phi$, and a model  whose $K^\mu$ does not have such a pole. Indeed, a closer inspection reveals not only how \eq{Am-symbol} might be deduced directly from \eq{Amu1} using heuristic arguments, but moreover, how \eq{Amu1}  could actually provide a numerical check of the model using \eq{Am-symbol}. In this regard, we note 
that the most straightforward way of adapting  \eq{Amu1}  to the case of models is simply to replace the exact bound state wave function with the approximate one of the model.
Yet in this way gauge invariance will be violated, unless the model BS kernel satisfies the consistency condition discussed in Appendix \ref{sec:consistency}. 
To preserve gauge invariance in the general case,  we first rewrite \eq{Amu1} for exact QCD in the form
\be
A^\mu_{ex} =e\gamma^\mu G_0^\mathit{ex} K_{ex}\Phi_{ex} \eqn{Amu1-it}
 \ee
and only then make the model approximations $K_{ex}\rightarrow K$ and $G_0^{ex}\rightarrow G_0$ in the equation for the bound state wave function $\Phi_{ex}$, and in the prefactor $e\gamma^\mu G_0^{ex} K_{ex}$ of \eq{Amu1-it}. The resulting approximation for the photon decay amplitude, $A^\mu$, will not be gauge invariant and will thus  differ from the exact amplitude  $A^\mu_{ex}$. A reasonable way to compensate for this difference is to add the 
term $S K^\mu$ to the prefactor, i.e., 
$e\gamma^\mu G_0^{ex} K_{ex}\rightarrow e\gamma^\mu G_0 K+S K^\mu$, as this  restores the gauge invariance of $A^\mu$. These replacements in \eq{Amu1-it}  lead to the gauge invariant expression of \eq{Am-symbol}:
\be
A^\mu =e\gamma^\mu G_0^{ex} K_{ex}\Phi_{ex}
\rightarrow (e\gamma^\mu G_0 K+S K^\mu)\Phi=(e\gamma^\mu +S K^\mu)\Phi . \eqn{Amu1-app}
 \ee
\begin{figure}[t]
\begin{fmffile}{Model}
\vspace{0.0cm}
\begin{align*} (a)& \ \ A^\mu_{ex} =  e\gamma^\mu G_0^{ex} \Biggl(\ \
\parbox{15mm}{
\begin{fmfgraph*}(15,10)
\fmfstraight
\fmfleftn{l}{2}\fmfrightn{r}{2}
\fmfset{curly_len}{2.mm}
\fmf{plain,tension=1.5}{r1,u,l1}
\fmf{plain,tension=1.5}{r2,v,l2}
\fmffreeze
\fmf{gluon}{v,u}
\end{fmfgraph*}} 
\ +\ 
\parbox{25mm}{
\begin{fmfgraph*}(25,10)
\fmfstraight
\fmfleftn{l}{2}\fmfrightn{r}{2}
\fmfset{curly_len}{2.mm}
\fmf{plain,tension=1.5}{r2,u2}
\fmf{plain,tension=.7}{u2,u1}
\fmf{plain,tension=1.5}{u1,l2}
\fmf{plain,tension=1.5}{l1,v1}
\fmf{plain,tension=.7}{v1,v2}
\fmf{plain,tension=1.5}{v2,r1}
\fmffreeze
\fmf{phantom}{u2,u,u1}
\fmf{phantom}{v2,v,v1}
\fmfv{d.s=circle,d.f=full,d.si=5}{u}
\fmfv{d.s=circle,d.f=full,d.si=5}{v}
\fmf{gluon}{v1,u2}
\fmf{gluon,rubout}{v2,u1}
\end{fmfgraph*}}
\ +\
\parbox{25mm}{
\begin{fmfgraph*}(25,10)
\fmfstraight
\fmfleftn{l}{2}\fmfrightn{r}{2}
\fmfset{curly_len}{2.mm}
\fmf{plain,tension=1.5}{r2,u2}
\fmf{plain,tension=1.2}{u2,uc,u1}
\fmf{plain,tension=1.5}{u1,l2}
\fmf{plain,tension=1}{l1,vc}
\fmf{plain,tension=1}{vc,r1}
\fmffreeze
\fmf{phantom}{u2,u,uc}
\fmf{phantom}{uc,v,u1}
\fmfv{d.s=circle,d.f=full,d.si=5}{u}
\fmfv{d.s=circle,d.f=full,d.si=5}{v}
\fmf{gluon,l.d=10}{vc,uc}
\fmf{gluon,right=.7}{u2,u1}
\end{fmfgraph*}} 
\ +\
\parbox{25mm}{
\begin{fmfgraph*}(25,10)
\fmfstraight
\fmfleftn{l}{2}\fmfrightn{r}{2}
\fmfset{curly_len}{2.mm}
\fmf{plain,tension=1}{r2,uc}
\fmf{plain,tension=1}{uc,l2}
\fmf{plain,tension=1.5}{l1,v1}
\fmf{plain,tension=1.2}{v1,vc,v2}
\fmf{plain,tension=1.5}{v2,r1}
\fmffreeze
\fmf{phantom}{v2,u,vc}
\fmf{phantom}{vc,v,v1}
\fmfv{d.s=circle,d.f=full,d.si=5}{u}
\fmfv{d.s=circle,d.f=full,d.si=5}{v}
\fmf{gluon}{vc,uc}
\fmf{gluon,right=.7}{v1,v2}
\end{fmfgraph*}} \ \  \Biggr)\Phi_{ex} \\[15mm]
 & = \ e\gamma^\mu G_0^{ex} \Biggl(\ \
\parbox{15mm}{
\begin{fmfgraph*}(15,10)
\fmfstraight
\fmfleftn{l}{2}\fmfrightn{r}{2}
\fmfset{curly_len}{2.mm}
\fmf{plain,tension=1.5}{r1,u,l1}
\fmf{plain,tension=1.5}{r2,v,l2}
\fmffreeze
\fmf{gluon}{v,u}
\end{fmfgraph*}} 
\ +\ 
\parbox{25mm}{
\begin{fmfgraph*}(25,10)
\fmfstraight
\fmfleftn{l}{2}\fmfrightn{r}{2}
\fmfset{curly_len}{2.mm}
\fmf{plain,tension=1.5}{r2,u2}
\fmf{plain,tension=.7}{u2,u1}
\fmf{plain,tension=1.5}{u1,l2}
\fmf{plain,tension=1.5}{l1,v1}
\fmf{plain,tension=.7}{v1,v2}
\fmf{plain,tension=1.5}{v2,r1}
\fmffreeze
\fmf{phantom}{u2,u,u1}
\fmf{phantom}{v2,v,v1}
\fmfv{d.s=circle,d.f=full,d.si=5}{u}
\fmfv{d.s=circle,d.f=full,d.si=5}{v}
\fmf{gluon}{v1,u2}
\fmf{gluon,rubout}{v2,u1}
\end{fmfgraph*}} \  \ \Biggr) \Phi_{ex}
\ + \ \Biggl(\ 
\parbox{25mm}{
\begin{fmfgraph*}(25,10)
\fmfstraight
\fmfleftn{l}{3}\fmfrightn{r}{3}
\fmfset{curly_len}{2.mm}
\fmf{plain,tension=1.5}{r3,u3}
\fmf{plain,tension=.7}{u3,u1}
\fmf{phantom,tension=1.5}{u1,l3}
\fmf{phantom,tension=1.5}{l1,v1}
\fmf{plain,tension=.7}{v1,v3}
\fmf{plain,tension=1.5}{v3,r1}
\fmffreeze
\fmf{plain,tension=1.5,right=0.45}{u1,l2}
\fmf{plain,tension=1.5,left=0.45}{v1,l2}
\fmf{phantom}{u3,u,u1}
\fmf{phantom}{v3,v,v1}
\fmf{phantom}{v3,vd,v1}
\fmfv{d.s=circle,d.f=full,d.si=5}{l2}
\fmfv{d.s=circle,d.f=full,d.si=5}{u}
\fmffreeze
\fmfshift{(0,-25)}{vd}
\fmffreeze
\fmf{photon}{v,vd}
\fmf{gluon}{v1,u3}
\fmf{gluon,rubout}{v3,u1}
\end{fmfgraph*}}
\ +\ 
\parbox{25mm}{
\begin{fmfgraph*}(25,10)
\fmfstraight
\fmfleftn{l}{3}\fmfrightn{r}{3}
\fmfset{curly_len}{2.mm}
\fmf{plain,tension=1.5}{r3,u3}
\fmf{plain,tension=.7}{u3,u1}
\fmf{phantom,tension=1.5}{u1,l3}
\fmf{phantom,tension=1.5}{l1,v1}
\fmf{plain,tension=.7}{v1,v3}
\fmf{plain,tension=1.5}{v3,r1}
\fmffreeze
\fmf{plain,tension=1.5,right=0.45}{u1,l2}
\fmf{plain,tension=1.5,left=0.45}{v1,l2}
\fmf{phantom}{u3,u,u1}
\fmf{phantom}{v3,v,v1}
\fmf{phantom}{u3,ud,u1}
\fmfv{d.s=circle,d.f=full,d.si=5}{l2}
\fmfv{d.s=circle,d.f=full,d.si=5}{v}
\fmffreeze
\fmfshift{(0,25)}{ud}
\fmffreeze
\fmf{photon}{u,ud}
\fmf{gluon}{v1,u3}
\fmf{gluon,rubout}{v3,u1}
\end{fmfgraph*}}\ \ \Biggr)\Phi_{ex}
 \\[15mm]
\rightarrow & \ \  A^\mu =  e\gamma^\mu G_0 \Biggl(\ \
\parbox{15mm}{
\begin{fmfgraph*}(15,10)
\fmfstraight
\fmfleftn{l}{2}\fmfrightn{r}{2}
\fmfset{curly_len}{2.mm}
\fmf{plain,tension=1.5}{r1,u,l1}
\fmf{plain,tension=1.5}{r2,v,l2}
\fmffreeze
\fmf{gluon}{v,u}
\end{fmfgraph*}} 
\ +\ 
\parbox{25mm}{
\begin{fmfgraph*}(25,10)
\fmfstraight
\fmfleftn{l}{2}\fmfrightn{r}{2}
\fmfset{curly_len}{2.mm}
\fmf{plain,tension=1.5}{r2,u2}
\fmf{plain,tension=.7}{u2,u1}
\fmf{plain,tension=1.5}{u1,l2}
\fmf{plain,tension=1.5}{l1,v1}
\fmf{plain,tension=.7}{v1,v2}
\fmf{plain,tension=1.5}{v2,r1}
\fmffreeze
\fmf{phantom}{u2,u,u1}
\fmf{phantom}{v2,v,v1}
\fmf{gluon}{v1,u2}
\fmf{gluon,rubout}{v2,u1}
\end{fmfgraph*}} \  \ \Biggr) \Phi
\ + \ \Biggl(\ 
\parbox{25mm}{
\begin{fmfgraph*}(25,10)
\fmfstraight
\fmfleftn{l}{3}\fmfrightn{r}{3}
\fmfset{curly_len}{2.mm}
\fmf{plain,tension=1.5}{r3,u3}
\fmf{plain,tension=.7}{u3,u1}
\fmf{phantom,tension=1.5}{u1,l3}
\fmf{phantom,tension=1.5}{l1,v1}
\fmf{plain,tension=.7}{v1,v3}
\fmf{plain,tension=1.5}{v3,r1}
\fmffreeze
\fmf{plain,tension=1.5,right=0.45}{u1,l2}
\fmf{plain,tension=1.5,left=0.45}{v1,l2}
\fmf{phantom}{u3,u,u1}
\fmf{phantom}{v3,v,v1}
\fmf{phantom}{v3,vd,v1}
\fmfv{d.s=circle,d.f=full,d.si=5}{l2}
\fmffreeze
\fmfshift{(0,-25)}{vd}
\fmffreeze
\fmf{photon}{v,vd}
\fmf{gluon}{v1,u3}
\fmf{gluon,rubout}{v3,u1}
\end{fmfgraph*}}
\ +\ 
\parbox{25mm}{
\begin{fmfgraph*}(25,10)
\fmfstraight
\fmfleftn{l}{3}\fmfrightn{r}{3}
\fmfset{curly_len}{2.mm}
\fmf{plain,tension=1.5}{r3,u3}
\fmf{plain,tension=.7}{u3,u1}
\fmf{phantom,tension=1.5}{u1,l3}
\fmf{phantom,tension=1.5}{l1,v1}
\fmf{plain,tension=.7}{v1,v3}
\fmf{plain,tension=1.5}{v3,r1}
\fmffreeze
\fmf{plain,tension=1.5,right=0.45}{u1,l2}
\fmf{plain,tension=1.5,left=0.45}{v1,l2}
\fmf{phantom}{u3,u,u1}
\fmf{phantom}{v3,v,v1}
\fmf{phantom}{u3,ud,u1}
\fmfv{d.s=circle,d.f=full,d.si=5}{l2}
\fmffreeze
\fmfshift{(0,25)}{ud}
\fmffreeze
\fmf{photon}{u,ud}
\fmf{gluon}{v1,u3}
\fmf{gluon,rubout}{v3,u1}
\end{fmfgraph*}}\ \ \Biggr)\Phi
 \\[15mm]
(b) & \ \ \Sigma_{ex}\ = \ 
\parbox{22mm}{
\begin{fmfgraph*}(22,10)
\fmfstraight
\fmfleftn{l}{1}\fmfrightn{r}{1}
\fmfset{curly_len}{1.8mm}
\fmf{plain,tension=1.}{r1,v2}
\fmf{plain,tension=0.5}{v2,v1}
\fmf{plain,tension=1.}{v1,l1}
\fmffreeze
\fmf{phantom}{v1,a,v2}
\fmfv{d.s=circle,d.f=full,d.si=5}{a}
\fmf{gluon,right=.85}{v2,v1}
\end{fmfgraph*}} 
\ + \
\parbox{35mm}{
\begin{fmfgraph*}(35,10)
\fmfstraight
\fmfleftn{l}{1}\fmfrightn{r}{1}
\fmfset{curly_len}{1.8mm}
\fmf{plain,tension=1}{r1,v4}
\fmf{plain,tension=.7}{v4,v3}
\fmf{plain,tension=.6}{v3,v2}
\fmf{plain,tension=.7}{v2,v1}
\fmf{plain,tension=1}{v1,l1}
\fmffreeze
\fmf{phantom}{v1,a,v2}
\fmf{phantom}{v2,b,v3}
\fmf{phantom}{v3,c,v4}
\fmfv{d.s=circle,d.f=full,d.si=5}{a}
\fmfv{d.s=circle,d.f=full,d.si=5}{b}
\fmfv{d.s=circle,d.f=full,d.si=5}{c}
\fmf{gluon,right=.7}{v3,v1}
\fmf{gluon,right=.7,rubout}{v4,v2}
\end{fmfgraph*}} 
\ \  \rightarrow \  \ \Sigma =\
\parbox{22mm}{
\begin{fmfgraph*}(22,10)
\fmfstraight
\fmfleftn{l}{1}\fmfrightn{r}{1}
\fmfset{curly_len}{1.8mm}
\fmf{plain,tension=1.}{r1,v2}
\fmf{plain,tension=0.5}{v2,v1}
\fmf{plain,tension=1.}{v1,l1}
\fmffreeze
\fmf{phantom}{v1,a,v2}
\fmfv{d.s=circle,d.f=full,d.si=5}{a}
\fmf{gluon,right=.85}{v2,v1}
\end{fmfgraph*}} 
\ + \
\parbox{35mm}{
\begin{fmfgraph*}(35,10)
\fmfstraight
\fmfleftn{l}{1}\fmfrightn{r}{1}
\fmfset{curly_len}{1.8mm}
\fmf{plain,tension=1}{r1,v4}
\fmf{plain,tension=.7}{v4,v3}
\fmf{plain,tension=.6}{v3,v2}
\fmf{plain,tension=.7}{v2,v1}
\fmf{plain,tension=1}{v1,l1}
\fmffreeze
\fmf{phantom}{v1,a,v2}
\fmf{phantom}{v2,b,v3}
\fmf{phantom}{v3,c,v4}
\fmfv{d.s=circle,d.f=full,d.si=5}{b}
\fmf{gluon,right=.7}{v3,v1}
\fmf{gluon,right=.7,rubout}{v4,v2}
\end{fmfgraph*}} 
\end{align*}
\end{fmffile}
\vspace{0mm}
\caption{\fign{MM} Illustration of how \eq{Amu1} for an "exact" theory becomes \eq{Am-symbol} for a "model" theory. (a) The one-photon decay amplitude in the exact theory, expressed as $A^\mu_{ex} = e\gamma^\mu \Phi_{ex}= e\gamma^\mu G_0^{ex} K_{ex} \Phi_{ex}$, where the exact kernel $K_{ex}$ consists of the 4 one- and two-gluon exchange contributions, as shown in the first line, is approximated by the model decay amplitude, given by $A^\mu = e\gamma^\mu G_0 K  \Phi + SK^\mu \Phi=(e\gamma^\mu  + SK^\mu) \Phi$, where the model kernel $K$ consists of the 2 one- and two-gluon exchange contributions, as shown in the third line. (b) The corresponding quark self-energy terms needed to ensure gauge invariance in both the exact and model theories.
 }
\end{figure}
What makes this heuristic argument particularly compelling is that it can be applied to specific exact descriptions, together with their models, thereby enabling a numerical consistency check between the exact result given by \eq{Amu1}  and the model result given by \eq{Am-symbol}. A transparent example of this is given in \fig{MM}. Here we start with an exact description where  the BS kernel $K_{ex}$ is chosen as the sum of 4 one- and two-gluon exchange diagrams with dressed intermediate-state quarks, as illustrated in the first line of \fig{MM}. In particular, the first two lines of \fig{MM} illustrate how the exact one-photon decay amplitude, as given by $A^\mu_{ex} = e\gamma^\mu \Phi_{ex}= e\gamma^\mu G_0^{ex} K_{ex} \Phi_{ex}$, can be split into two parts, one containing just the one- and crossed-two-gluon exchange contributions to $ K_{ex} $, and the other containing the vertex dressing contributions (note that in the latter case  $e\gamma^\mu$ is illustrated as a bare quark-photon vertex). The third line of \fig{MM} then shows how the exact description is replaced with a model one; namely, all but two of the intermediate-state exact dressed quark propagators within each diagram of  $K_{ex}$ are replaced by bare propagators of physical mass, and the  exact $\Phi$-meson bound state wave function $\Phi_{ex}$ is replaced by a model wave function $\Phi$ whose BS kernel $K$ is given by the sum of 2 one- and two-crossed-gluon exchange diagrams. To specify how the exact free Green function $G_0^{ex}$ is replaced by the model free Green function $G_0$, we refer to \fig{MM}(b) where it is shown how the corresponding exact quark self-dressing term $\Sigma_{ex}$ is replaced by the model $\Sigma$ (note that, as expected, $\Sigma^\mu_{ex} = S_{ex}^\mu K_{ex}$ and $\Sigma = S K$).
In this way one obtains that the model decay amplitude is given by the expression $A^\mu = e\gamma^\mu G_0 K \Phi + S K^\mu \Phi = (e\gamma^\mu  + S K^\mu) \Phi$.
Assuming that the quark-gluon vertices are renormalized, the contribution of the vertex dressing diagrams (last two diagrams in the first line of \fig{MM}) is expected to be small. Similarly, the replacement of intermediate-state dressed quark propagators by bare ones (but of physical mass) is expected to give only a small effect. Thus, overall, one would expect that $A_{ex}^\mu \approx A^\mu$ numerically, yet they are given by two different gauge invariant expressions, \eq{Amu1}  and \eq{Am-symbol}, respectively.
Writing the symbolic-notation \eq{Am-symbol} out in full,
 \begin{align}
A^\mu &=\int \frac{d^4p}{(2\pi)^4} \mbox{Tr}[\bar F^\mu(p)\Phi(P,p)] \eqn{Amn} \\
 &=\int \frac{d^4p\, d^4k}{(2\pi)^8} \bigl\{[e\gamma^\mu]^a_b+S^c_d(k) [K^\mu]^{da}_{cb}(k,k;p,p')\bigr\}\Phi^b_a(P,p) \eqn{1} 
\end{align}
where the model bound state wave function $\Phi(P,p)$ is determined by the BS equation, \eq{BS}. Equation (\ref{1}) is illustrated in \fig{fig:Am}.
\begin{figure}[t]
\begin{fmffile}{Amu}
\begin{align*}
A^\mu\ \ =\quad
\parbox{35mm}{
\begin{fmfgraph*}(35,30)
\fmfstraight
\fmfleft{l}\fmfright{r}
\fmf{photon,tension=1,l.s=left,label=$q\rightarrow$}{l,v}
\fmf{fermion,tension=.4,right,l.s=right,label=$p,,a$}{phi,v}
\fmf{fermion,tension=.4,right,l.s=right,label=$p',,b$}{v,phi}
\fmf{phantom,tension=1.8}{phi,r}
\fmffreeze
\fmfv{d.sh=circle,d.f=empty,d.si=18,l=$\hspace{-3.2mm}\Phi$}{phi}
\fmfi{plain}{vpath (__phi,__r) shifted (thick*(0,1.))}
\fmfi{plain}{vpath (__phi,__r) shifted (thick*(0,-1.))}
\end{fmfgraph*}} \quad
+\quad
\parbox{50mm}{
\begin{fmfgraph*}(50,30)
\fmfstraight
\fmfright{r}\fmfleft{l}
\fmf{phantom,tension=1.5}{l,v4}
\fmf{phantom,tension=1}{v4,v2}
\fmf{phantom,tension=1}{v2,phi}
\fmf{phantom,tension=2.4}{phi,r}
\fmffreeze
\fmf{phantom,tension=25}{v3,v2,v1}
\fmf{phantom,tension=1}{l,v3}
\fmf{phantom,tension=1}{v1,r}
\fmffreeze
\fmf{fermion,tension=.4,right,l.s=right,label=$p,,a$}{phi,v1}
\fmf{fermion,tension=.4,right,l.s=right,label=$p',,b$}{v1,phi}
\fmf{fermion,tension=.4,right=1,label=$k,,c$}{v3,v4}
\fmf{fermion,tension=.4,right=1,label=$k,,d$}{v4,v3}
\fmfv{d.s=circle,d.f=full,d.si=7}{v4}
\fmf{photon, left=.3,rubout=4,l.s=left,l.d=3,label=$\hspace{-1.5cm}q\rightarrow$}{l,v3}
\fmffreeze
\fmfv{d.sh=circle,d.f=empty,d.si=18,l=$\hspace{-3.2mm}\Phi$}{phi}
\fmfi{plain}{vpath (__phi,__r) shifted (thick*(0,1.))}
\fmfi{plain}{vpath (__phi,__r) shifted (thick*(0,-1.))}
\fmfv{d.s=square,d.f=full,d.si=8}{v2}
\end{fmfgraph*}}
\end{align*}
\end{fmffile} 
\vspace{0mm}

\caption{\fign{fig:Am} Illustration of the one-photon decay amplitude $A^\mu$ as given by \eq{1}.}
\end{figure}
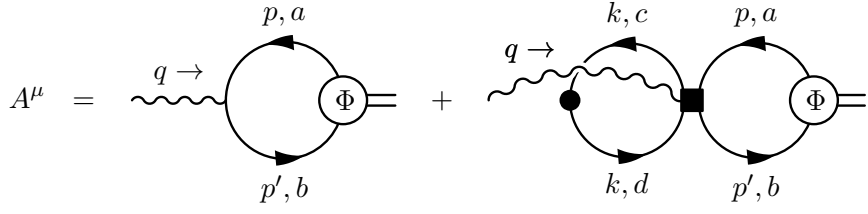
The gauge invariance of the one-photon decay amplitude, as given by \eq{1}, rests crucially on the assumption that all quantities in the equation are derived from the same BS kernel. Thus the BS kernel $K$ that determines the dressed quark propagator $S$ through the DS equation (\ref{DS}), is the same kernel that determines the wave function $\Phi$ through the BS equation (\ref{BS}). In addition, the  five-point function $K^\mu$ needs to be constructed by attaching photons to all places in the same $K$ (by gauging the expression for $K$), thus guaranteeing that $K^\mu$ satisfies the WT identity\footnote{Here the indices $a_1, b_1, a_2, b_2$ and $c_1$ represent just the isospin components.}
\begin{align}\label{WTI-K}
q_\mu [K^\mu]&^{b_2a_1}_{b_1a_2} (k_1,k_2;p_1,p_2)\nn
&=
i[e^{c_1}_{b_1}K^{b_2a_1}_{c_1a_2}(k_1-q,k_2;p_1,p_2)-
e_{c_2}^{b_2}K^{c_2a_1}_{b_1a_2}(k_1,k_2+q;p_1,p_2)
\nn
&
-K^{b_2c_1}_{b_1a_2}(k_1,k_2;p_1+q,p_2)e^{a_1}_{c_1}+K^{b_2a_1}_{b_1c_2}(k_1,k_2;p_1,p_2-q)e_{a_2}^{c_2}].
\end{align}
Note that $K^\mu$ is related to the five-point Green function $G^\mu$ which in QFT is defined by
\begin{align}
 G^\mu(k_1,k_2;p_1,p_2)
 &=\int d^4y_1\,d^4y_2\,d^4x_1\,d^4x_2\, e^{i(k_1y_1-k_2y_2-p_1x_1+p_2x_2)}
\nn
&\hspace{1cm}\times \la0|Tq(y_1)\bar q(y_2)\bar q(x_1)q(x_2)J^\mu(0)|0\ra;
\end{align}
in particular \cite{KB3},
\be
G^\mu = G \left(\Gamma_0^\mu + K^\mu\right) G   \eqn{Gammamu-2b}
\ee
where
\be
\Gamma_0^\mu = G_0^{-1} G_0^\mu G_0^{-1}
\ee
is the sum of the constituent currents (further discussed in Appendix B).
Equation (\ref{WTI-K}) and its analog for $G^\mu$ can be written in the symbolic form
\begin{subequations}\eqn{WT-K2r-pa}
\begin{align}
q_\mu K^\mu &= [\hat e,K],\\
q_\mu G^\mu &=[ \hat e,G] , \eqn{WTG}
\end{align}
\end{subequations}
where the operator $ \hat e$ is defined as 
\be
 \hat e^{b_2a_1}_{b_1a_2}(k_1,k_2;p_1,p_2)
=i(2\pi)^4\left[e^{a_1}_{b_1}\delta^{b_2}_{a_2}\delta^4(k_2-p_2)-e^{b_2}_{a_2}\delta^{a_1}_{b_1}\delta^4(k_1-p_1)\right]  \eqn{hat-e}
\ee
where $k_1-k_2-q=p_1-p_2$ because the total momentum conservation delta function 
$(2\pi)^4\delta^4(k_1-p_1-k_2+p_2-q)$, is factored out.
A similar operator was introduced in Eq.\ (103) of Ref.\ \cite{KB3} acting in the space of three quark states. The minus sign between delta functions in \eq{hat-e} is due to the fact that the operator $\hat e$ acts in the space of $q\bar q$ states. 
Because the one-photon decay amplitude $A^\mu$ results from taking a residue at the bound state pole of the three-point Green function $S^\mu$, its gauge invariance is expressed through the WT identity for $S^\mu$. Thus, to complete the description of the one-photon decay amplitude, as given by the set of equations (\ref{bFmu-full} - \ref{WT-K2r-pa}), one should add the equation which determines $S^\mu$ using the same BS kernel $K$:
\be
 S^\mu(p',p)=S(p') F^\mu(p) S(p)+\int \frac{d^4k}{(2\pi)^4}  [G_0K](p',p;k',k) S^\mu(k) \eqn{DSmu}
\ee
which follows from \eqs{Gammamu} and (\ref{Gammamudef}). Because \eq{DSmu} results from gauging the DS equation, \eq{DS}, its solution $S^\mu$ will satisfy the WT identity 
\be
q_\mu S^\mu(p',p) =i\left[e S(p)-S(p')e\right].   \eqn{wti}
\ee
Equation (\ref{DSmu}) is a necessary ingredient  for the formulation of the two-photon decay amplitude where gauge invariance requires attachment of photons to dressed quarks [see \eq{2gam-symb}].

\subsection{Gauge invariant \bm{$\Phi\rightarrow 2\gamma$} amplitude}
\label{phi-2g}

In QFT, the two-photon decay amplitude $A^{\mu\nu}$ is defined by
\be
A^{\mu\nu}=\int d^4x\,e^{-iq_2x}\la0|J^\nu(x)J^\mu(0)|\Phi\ra \eqn{Amndef}
\ee
where the current operators $J^\mu$ and $J^\nu$ are associated with incoming photons of momenta and polarisation indices $(q_1, \mu)$ and $(q_2, \nu)$, respectively.
From a diagrammatic point of view, $A^{\mu\nu}$ contains all of the diagrams that result from attaching photons of polarisation index $\nu$ to all possible places in the Feynman diagrams contributing to the one-photon decay amplitude $A^\mu=\la0|J^\mu(0)|\Phi\ra$. As such, our $A^{\mu\nu}$ amplitude is derived by gauging the expression for $A^\mu$ given in \eq{Amn}. The details of this gauging are presented in Appndix B. Alternatively, one could gauge the dressed quark propagator twice, $S\rightarrow S^\mu\rightarrow S^{\mu\nu}$,  and then extract the $2\gamma$ decay amplitude from the pole in the $q\bar q$ channel corresponding to a meson bound state; however, the formal procedure of gauging  \eq{Amn} leads to the same result.   Given this formal equivalence of the two methods, we can ignore the fact that selection rules may allow one of the processes $\Phi\rightarrow\gamma$ or $\Phi\rightarrow 2\gamma$, but forbid the other. We can also ignore that the effective mass of the virtual photon in  \eq{Amn} is equal to the mass of the meson, and consider it arbitrary in the expressions for two- and three-photon decays. In this way we obtain the following expression for $A^{\mu\nu}$:
\begin{align}\label{2gam-symb}
A^{\mu\nu}=\mbox{Tr}\, \left(\Gamma^\mu S \Gamma^\nu +\Gamma^\nu S \Gamma^\mu\right) \Phi
+
\left(S^\mu K^\nu+S^\nu K^\mu+S K^{\mu\nu}\right)\Phi.
\end{align}
The full expression symbolised by \eq{2gam-symb} is
\begin{align}
A^{\mu\nu}&=
\mbox{Tr}\int \frac{d^4p}{(2\pi)^4} \Gamma^\mu(p+q_2)S(p+q_2)\Gamma^\nu(p)\Phi(P,p)
\nn
&+\mbox{Tr}\int \frac{d^4p}{(2\pi)^4}\Gamma^\nu(p+q_1)S(p+q_1)\Gamma^\mu(p)\Phi(P,p)
\nn[2mm]
&+\int \frac{d^4p\,d^4k}{(2\pi)^8} \bigl\{ [S^\mu]^c_d(k) [K^\nu]^{da}_{cb}(k,k+q_1;p,p+q_1+q_2) \nn
&\hspace{2.8cm} +[S^\nu]^c_d(k)[K^\mu]^{da}_{cb}(k,k+q_2;p,p+q_1+q_2)
\bigr\} \Phi^b_a(P,p)\nn[2mm]
&+\int \frac{d^4p\,d^4k}{(2\pi)^8} S^c_d(k)
[K^{\mu\nu}]^{da}_{cb}(k,k;p,p+q_1+q_2) 
\Phi^b_a(P,p)      \eqn{2gam-full}
\end{align}
where the superscripts $\mu$ and $\nu$ on gauged quantities give them implicit dependence on the photon momenta $q_1$ and $q_2$, respectively. Equation (\ref{2gam-full}) is illustrated in \fig{fig:Amn}.
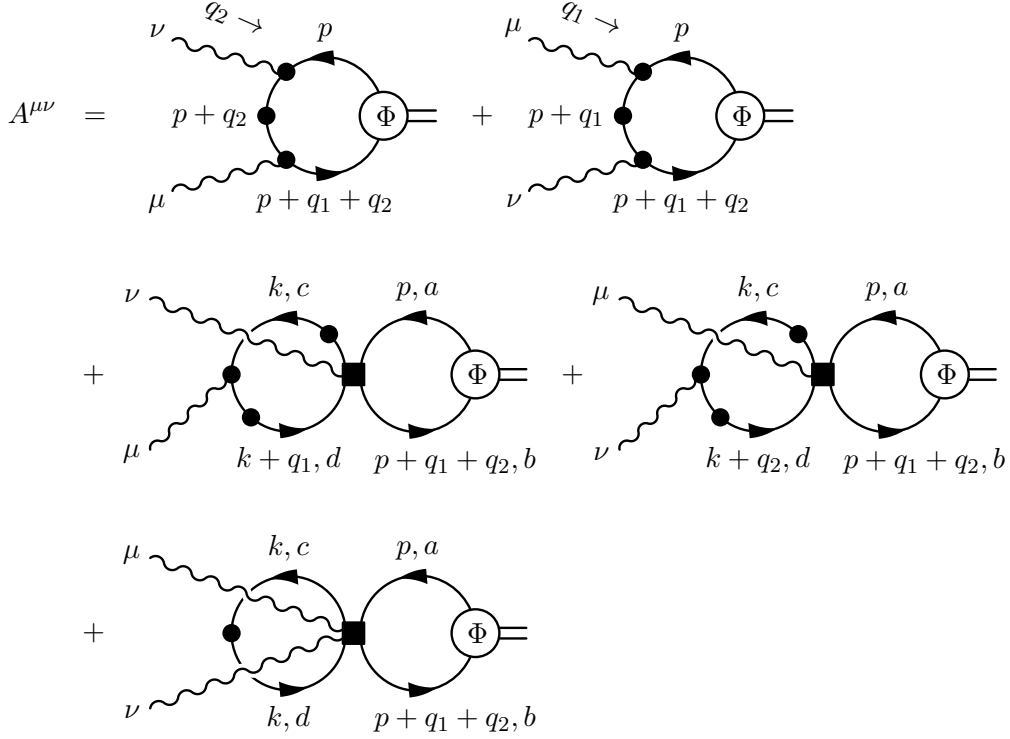
\begin{figure}[t]
\begin{fmffile}{Amn}
\begin{align*}
A^{\mu\nu}\ \ &=\quad\quad
\parbox{35mm}{
\begin{fmfgraph*}(35,20)
\fmfstraight
\fmfleftn{l}{3}\fmfrightn{r}{3}
\fmf{phantom,tension=1}{l2,v}
\fmf{fermion,tension=.4,right,l.s=right,label=$p$}{phi,v}
\fmf{fermion,tension=.4,right,l.s=right,label=$p+q_1+q_2$}{v,phi}
\fmf{phantom,tension=1.8}{phi,r2}
\fmffreeze
\fmfv{d.s=circle,d.f=full,d.si=6}{v}
\fmfv{d.sh=circle,d.f=empty,d.si=18,l=$\hspace{-3.2mm}\Phi$}{phi}
\fmf{phantom,left,tag=1}{v,phi}
\fmf{phantom,right,tag=2}{v,phi}
\fmfposition
\fmfipath{p[]}
\fmfiset{p1}{vpath1(__v,__phi)}
\fmfiset{p2}{vpath2(__v,__phi)}
\fmfi{photon,label=\rotatebox{-24}{$q_2\rightarrow$}}{point length(p1)/4 of p1 -- vloc(__l3)}
\fmfv{l.d=3,label=$\nu$}{l3}
\fmfv{l.d=3,label=$\mu$}{l1}
\fmfv{l.d=4,l.d=7,label=$p+q_2$}{v}
\fmfi{photon}{point length(p2)/4 of p2 -- vloc(__l1)}
\fmfiv{d.s=circle,d.f=full,d.si=6}{point length(p1)/4 of p1}
\fmfiv{d.s=circle,d.f=full,d.si=6}{point length(p2)/4 of p2}
\fmfi{plain}{vpath (__phi,__r2) shifted (thick*(0,1.))}
\fmfi{plain}{vpath (__phi,__r2) shifted (thick*(0,-1.))}
\end{fmfgraph*}} \quad
+\quad
\parbox{35mm}{
\begin{fmfgraph*}(35,20)
\fmfstraight
\fmfleftn{l}{3}\fmfrightn{r}{3}
\fmf{phantom,tension=1}{l2,v}
\fmf{fermion,tension=.4,right,l.s=right,label=$p$}{phi,v}
\fmf{fermion,tension=.4,right,l.s=right,label=$p+q_1+q_2$}{v,phi}
\fmf{phantom,tension=1.8}{phi,r2}
\fmffreeze
\fmfv{d.s=circle,d.f=full,d.si=6}{v}
\fmfv{d.sh=circle,d.f=empty,d.si=18,l=$\hspace{-3.2mm}\Phi$}{phi}
\fmf{phantom,left,tag=1}{v,phi}
\fmf{phantom,right,tag=2}{v,phi}
\fmfposition
\fmfipath{p[]}
\fmfiset{p1}{vpath1(__v,__phi)}
\fmfiset{p2}{vpath2(__v,__phi)}
\fmfi{photon,label=\rotatebox{-24}{$q_1\rightarrow$}}{point length(p1)/4 of p1 -- vloc(__l3)}
\fmfi{photon,right}{point length(p2)/4 of p2 -- vloc(__l1)}
\fmfv{l.d=3,label=$\mu$}{l3}
\fmfv{l.d=3,label=$\nu$}{l1}
\fmfv{l.d=4,l.d=7,label=$p+q_1$}{v}
\fmfiv{d.s=circle,d.f=full,d.si=6}{point length(p1)/4 of p1}
\fmfiv{d.s=circle,d.f=full,d.si=6}{point length(p2)/4 of p2}
\fmfi{plain}{vpath (__phi,__r2) shifted (thick*(0,1.))}
\fmfi{plain}{vpath (__phi,__r2) shifted (thick*(0,-1.))}
\end{fmfgraph*}} \\[13mm]
\quad
&+\quad\
\parbox{50mm}{
\begin{fmfgraph*}(50,20)
\fmfstraight
\fmfrightn{r}{3}\fmfleftn{l}{3}
\fmf{phantom,tension=1.5}{l2,v4}
\fmf{phantom,tension=1}{v4,v2}
\fmf{phantom,tension=1}{v2,phi}
\fmf{phantom,tension=2.4}{phi,r2}
\fmffreeze
\fmf{phantom,tension=25}{v3,v2,v1}
\fmf{phantom,tension=1}{l2,v3}
\fmf{phantom,tension=1}{v1,r2}
\fmffreeze
\fmf{fermion,tension=.4,right,l.s=right,label=$p,,a$}{phi,v1}
\fmf{fermion,tension=.4,right,l.s=right,label=$\hspace{1cm}p+q_1+q_2,,b$}{v1,phi}
\fmf{fermion,tension=.4,right=1,label=$k,,c$}{v3,v4}
\fmf{fermion,tension=.4,right=1,label=$k+q_1,,d$}{v4,v3}
\fmfv{d.s=circle,d.f=full,d.si=6}{v4}
\fmf{photon,rubout=4}{l3,v3}
\fmf{photon}{l1,v4}
\fmfv{l.d=4,label=$\nu$}{l3}
\fmfv{l.d=4,label=$\mu$}{l1}
\fmffreeze
\fmf{phantom,right,tag=1}{v3,v4}
\fmf{phantom,left,tag=2}{v3,v4}
\fmfposition
\fmfipath{p[]}
\fmfiset{p1}{vpath1(__v3,__v4)}
\fmfiset{p2}{vpath2(__v3,__v4)}
\fmfiv{d.s=circle,d.f=full,d.si=6}{point length(p1)/4 of p1}
\fmfiv{d.s=circle,d.f=full,d.si=6}{point 3*length(p2)/4 of p2}
\fmfv{d.sh=circle,d.f=empty,d.si=18,l=$\hspace{-3.2mm}\Phi$}{phi}
\fmfi{plain}{vpath (__phi,__r2) shifted (thick*(0,1.))}
\fmfi{plain}{vpath (__phi,__r2) shifted (thick*(0,-1.))}
\fmfv{d.s=square,d.f=full,d.si=8}{v2}
\end{fmfgraph*}}\quad
+\quad
\parbox{50mm}{
\begin{fmfgraph*}(50,20)
\fmfstraight
\fmfrightn{r}{3}\fmfleftn{l}{3}
\fmf{phantom,tension=1.5}{l2,v4}
\fmf{phantom,tension=1}{v4,v2}
\fmf{phantom,tension=1}{v2,phi}
\fmf{phantom,tension=2.4}{phi,r2}
\fmffreeze
\fmf{phantom,tension=25}{v3,v2,v1}
\fmf{phantom,tension=1}{l2,v3}
\fmf{phantom,tension=1}{v1,r2}
\fmffreeze
\fmf{fermion,tension=.4,right,l.s=right,label=$p,,a$}{phi,v1}
\fmf{fermion,tension=.4,right,l.s=right,label=$\hspace{1cm}p+q_1+q_2,,b$}{v1,phi}
\fmf{fermion,tension=.4,right=1,label=$k,,c$}{v3,v4}
\fmf{fermion,tension=.4,right=1,label=$k+q_2,,d$}{v4,v3}
\fmfv{d.s=circle,d.f=full,d.si=6}{v4}
\fmf{photon,rubout=4}{l3,v3}
\fmf{photon}{l1,v4}
\fmfv{l.d=4,label=$\mu$}{l3}
\fmfv{l.d=4,label=$\nu$}{l1}
\fmffreeze
\fmf{phantom,right,tag=1}{v3,v4}
\fmf{phantom,left,tag=2}{v3,v4}
\fmfposition
\fmfipath{p[]}
\fmfiset{p1}{vpath1(__v3,__v4)}
\fmfiset{p2}{vpath2(__v3,__v4)}
\fmfiv{d.s=circle,d.f=full,d.si=6}{point length(p1)/4 of p1}
\fmfiv{d.s=circle,d.f=full,d.si=6}{point 3*length(p2)/4 of p2}
\fmfv{d.sh=circle,d.f=empty,d.si=18,l=$\hspace{-3.2mm}\Phi$}{phi}
\fmfi{plain}{vpath (__phi,__r2) shifted (thick*(0,1.))}
\fmfi{plain}{vpath (__phi,__r2) shifted (thick*(0,-1.))}
\fmfv{d.s=square,d.f=full,d.si=8}{v2}
\end{fmfgraph*}} \\[13mm]
&+\quad\
\parbox{50mm}{
\begin{fmfgraph*}(50,20)
\fmfstraight
\fmfrightn{r}{3}\fmfleftn{l}{3}
\fmf{phantom,tension=1.5}{l2,v4}
\fmf{phantom,tension=1}{v4,v2}
\fmf{phantom,tension=1}{v2,phi}
\fmf{phantom,tension=2.4}{phi,r2}
\fmffreeze
\fmf{phantom,tension=25}{v3,v2,v1}
\fmf{phantom,tension=1}{l2,v3}
\fmf{phantom,tension=1}{v1,r2}
\fmffreeze
\fmf{fermion,tension=.4,right,l.s=right,label=$p,,a$}{phi,v1}
\fmf{fermion,tension=.4,right,l.s=right,label=$\hspace{1cm}p+q_1+q_2,,b$}{v1,phi}
\fmf{fermion,tension=.4,right=1,label=$k,,c$}{v3,v4}
\fmf{fermion,tension=.4,right=1,label=$k,,d$}{v4,v3}
\fmfv{d.s=circle,d.f=full,d.si=6}{v4}
\fmf{photon,rubout=4}{v3,l1}
\fmf{photon,rubout=4}{v2,l3}
\fmfv{l.d=4,label=$\mu$}{l3}
\fmfv{l.d=4,label=$\nu$}{l1}
\fmffreeze
\fmfv{d.sh=circle,d.f=empty,d.si=18,l=$\hspace{-3.2mm}\Phi$}{phi}
\fmfi{plain}{vpath (__phi,__r2) shifted (thick*(0,1.))}
\fmfi{plain}{vpath (__phi,__r2) shifted (thick*(0,-1.))}
\fmfv{d.s=square,d.f=full,d.si=8}{v2}
\end{fmfgraph*}}
\end{align*}
\end{fmffile} 
\vspace{0mm}

\caption{\fign{fig:Amn} Illustration of the two-photon decay amplitude $A^{\mu\nu}$ as given by \eq{2gam-full}.}
\end{figure}

In considering the gauge invariance of the construction presented here, it is useful to recall the discussion in Appendix A where we stressed that gauge invariance provides a stronger restriction on photon decay amplitudes than the more commonly used requirement of current conservation.  For two-photon decay, gauge invariance is expressed by the requirement that the corresponding four-point Green function $S^{\mu\nu}$, defined in QFT as
\be
S^{\mu\nu}(p) =
\int d^4x\,d^4y\,e^{-i(q_1x+q_2y+pz)}\la0|J^\mu(x)J^\nu(y)q(0) \q(z)|0\ra,
\ee
satisfy the WT identity
\be
q_2{}_\nu S^{\mu\nu}(p)=ie[S^\mu(p)-S^\mu(p+q_2)].  \eqn{Smn-Sm}
\ee
For the specific two-photon decay amplitude of \eq{2gam-full}, gauge invariance 
is respected because the functions $S(p)$, $S^\mu(p)$ and $\Phi(P,p)$ are chosen in a consistent way where their three corresponding equations,  \eqs{DS},  (\ref{DSmu}) and (\ref{BS}), are determined by
a universal kernel $K$. Using these equations, one can show that the expression for $S^{\mu\nu}$ 
constructed in the same approach as the meson decay amplitude (\ref{2gam-full}) [given explicitly by  \eq{Smunu}], satisfies the WT identity of \eq{Smn-Sm}.
The transversality of the two-photon decay amplitude of \eq{2gam-full} then follows as a consequence of the WT identity for $S^{\mu\nu}$:
\be
q_1{}_\mu A^{\mu\nu}=q_2{}_\nu A^{\mu\nu}=0 .  \eqn{Amn-0}
\ee
Although  the WT identity (\ref{Smn-Sm}) is obvious by construction [because the $S^{\mu\nu}$ of  \eq{Smunu} is obtained by attaching a photon of polarisation index $\nu$ everywhere in $S^\mu$], it is instructive to derive it explicitly. This is done in Appendix \ref{WTI}. 

Current conservation analogous to \eq{Amn-0} is derived in Eqs.\ (16)-(18) of Ref.\ \cite{Hanhart:2007wa} for the case where the kernel does not support charge flow, so that $[K,\hat e]=0$ and $K^\mu=0$. As emphasized  above, such a result is less restrictive than the gauge invariance relation of \eq{Smn-Sm}. Finally, it should be noted that a similar analysis has been recently presented, but in the context of Compton scattering \cite{Eichmann:2012mp}.

\subsection{Gauge invariant \bm{$\Phi\rightarrow 3\gamma$} amplitude}\label{phi-3g}

In QFT, the three-photon decay amplitude $A^{\mu\nu\sigma}$ is defined by
\be
A^{\mu\nu\sigma} = \int d^4y\, d^4x\, e^{-i(q_3y+q_2x)}\la0|TJ^\sigma(y)J^\nu(x)J^\mu(0)|\Phi\ra  \eqn{Amnsigdef}
\ee
where the current operators $J^\mu$, $J^\nu$ and $J^\sigma$ are associated with incoming photons of momenta and polarisation indices $(q_1, \mu)$,  $(q_2, \nu)$ and $(q_3, \sigma)$, respectively.
In Appendix  \ref{3gamma} we derive the gauge invariant expression for $A^{\mu\nu\sigma}$ by gauging \eq{2gam-symb} for the two-photon decay amplitude $A^{\mu\nu}$. The result we obtain has the symbolic form
\begin{align}
A^{\mu\nu\sigma} &=\sum_{perm}\mbox{Tr}\,  \Gamma^{\mu} S\Gamma^\nu S \Gamma^\sigma \Phi
+\sum_{cycle}\mbox{Tr}\left( \Gamma^{\mu\nu} S\Gamma^\sigma + \Gamma^\mu S \Gamma^{\nu\sigma}\right)\Phi\nn
&+ 
\sum_{perm} S \Gamma^\mu S\Gamma^\nu S K^\sigma \Phi
+ 
\sum_{cycle} (S \Gamma^{\mu\nu} S K^\sigma + S\Gamma^\mu S K^{\nu\sigma})\Phi + S K^{\mu\nu\sigma }\Phi . \eqn{Amns-symb}
\end{align}
where $\sum_{perm}$ and $\sum_{cycle}$ indicate sums over the six permutations and the three cyclic permutations of the photon polarization indices $(\mu,\nu,\sigma)$, respectively. Such permutations  are accompanied by corresponding permutations of the photon momenta ($q_1,q_2,q_3$). Equation (\ref{Amns-symb}) is illustrated in \fig{fig:Amns}.

\begin{figure}[t]
\begin{fmffile}{Amns}
\begin{align*}
A^{\mu\nu\sigma}\ \ &=\ \  \sum_{\underset{(\mu, \nu, \sigma)}{perm}}\quad\quad
\parbox{35mm}{
\begin{fmfgraph*}(35,20)
\fmfleftn{l}{3}\fmfrightn{r}{3}
\fmf{phantom,tension=1}{l2,v}
\fmf{plain,tension=.4,right}{phi,v}
\fmf{plain,tension=.4,right}{v,phi}
\fmf{phantom,tension=1.8}{phi,r2}
\fmffreeze
\fmfv{d.sh=circle,d.f=empty,d.si=18,l=$\hspace{-3.2mm}\Phi$}{phi}
\fmf{phantom,left,tag=1}{v,phi}
\fmf{phantom,right,tag=2}{v,phi}
\fmfposition
\fmfipath{p[]}
\fmfiset{p1}{vpath1(__v,__phi)}
\fmfiset{p2}{vpath2(__v,__phi)}
\fmfi{photon,l.s=right,l.d=4,label=\rotatebox{-15}{$q_3\rightarrow$}}{point 0.42*length(p1) of p1 -- vloc(__l3)}
\fmfv{l.d=3,label=$\sigma$}{l3}
\fmfv{l.d=3,label=$\nu$}{l2}
\fmfv{l.d=3,label=$\mu$}{l1}
\fmfi{photon,l.s=left,l.d=4,label=\rotatebox{15}{$q_1\rightarrow$}}{point 0.40*length(p2) of p2 -- vloc(__l1)}
\fmf{photon,l.d=5,label=\rotatebox{-4}{$q_2\rightarrow$}}{v,l2}
\fmfiv{d.s=circle,d.f=full,d.si=6}{point 0.40*length(p1) of p1}
\fmfiv{d.s=circle,d.f=full,d.si=6}{point 0.40*length(p2) of p2}
\fmfiv{d.s=circle,d.f=full,d.si=6}{point 0.40*length(p1)/2 of p1}
\fmfiv{d.s=circle,d.f=full,d.si=6}{point 0.40*length(p2)/2 of p2}
\fmfv{d.s=circle,d.f=full,d.si=6}{v}
\fmfi{fermion,l.s=right,label=$p$}{subpath (length(p1),length(p1)/2) of p1}
\fmfi{fermion,l.s=right,label=$p'$}{subpath (length(p2)/2,length(p2)) of p2}
\fmfi{plain}{vpath (__phi,__r2) shifted (thick*(0,1.))}
\fmfi{plain}{vpath (__phi,__r2) shifted (thick*(0,-1.))}
\end{fmfgraph*}} \\[13mm]
 \quad
 &+\ \  \sum_{\underset{(\mu, \nu, \sigma)}{cyclic}}\quad\quad
\parbox{35mm}{
\begin{fmfgraph*}(35,20)
\fmfleftn{l}{3}\fmfrightn{r}{3}
\fmf{phantom,tension=1}{l2,v}
\fmf{fermion,tension=.4,right,l.s=right,label=$p$}{phi,v}
\fmf{fermion,tension=.4,right,l.s=right,label=$p'$}{v,phi}
\fmf{phantom,tension=1.8}{phi,r2}
\fmffreeze
\fmfv{d.sh=circle,d.f=empty,d.si=18,l=$\hspace{-3.2mm}\Phi$}{phi}
\fmf{phantom,left,tag=1}{v,phi}
\fmf{phantom,right,tag=2}{v,phi}
\fmfposition
\fmfipath{p[]}
\fmfiset{p1}{vpath1(__v,__phi)}
\fmfiset{p2}{vpath2(__v,__phi)}
\fmfi{photon}{point length(p1)/4 of p1 -- vloc(__l3)}
\fmfi{photon}{point length(p2)/4 of p2 -- vloc(__l2)}
\fmfi{photon,right}{point length(p2)/4 of p2 -- vloc(__l1)}
\fmfv{l.d=3,label=$\sigma$}{l3}
\fmfv{l.d=3,label=$\nu$}{l2}
\fmfv{l.d=3,label=$\mu$}{l1}
\fmfiv{d.s=circle,d.f=full,d.si=6}{point length(p1)/4 of p1}
\fmfiv{d.s=circle,d.f=full,d.si=6}{point length(p2)/4 of p2}
\fmfiv{d.s=circle,d.f=full,d.si=6}{point 0*length(p2)/4 of p2}
\fmfi{plain}{vpath (__phi,__r2) shifted (thick*(0,1.))}
\fmfi{plain}{vpath (__phi,__r2) shifted (thick*(0,-1.))}
\end{fmfgraph*}} \quad
+\ \  \sum_{\underset{(\mu, \nu, \sigma)}{cyclic}}\quad\quad
\parbox{35mm}{
\begin{fmfgraph*}(35,20)
\fmfleftn{l}{3}\fmfrightn{r}{3}
\fmf{phantom,tension=1}{l2,v}
\fmf{fermion,tension=.4,right,l.s=right,label=$p$}{phi,v}
\fmf{fermion,tension=.4,right,l.s=right,label=$p'$}{v,phi}
\fmf{phantom,tension=1.8}{phi,r2}
\fmffreeze
\fmfv{d.sh=circle,d.f=empty,d.si=18,l=$\hspace{-3.2mm}\Phi$}{phi}
\fmf{phantom,left,tag=1}{v,phi}
\fmf{phantom,right,tag=2}{v,phi}
\fmfposition
\fmfipath{p[]}
\fmfiset{p1}{vpath1(__v,__phi)}
\fmfiset{p2}{vpath2(__v,__phi)}
\fmfi{photon}{point length(p1)/4 of p1 -- vloc(__l3)}
\fmfi{photon}{point length(p1)/4 of p1 -- vloc(__l2)}
\fmfi{photon,right}{point length(p2)/4 of p2 -- vloc(__l1)}
\fmfv{l.d=3,label=$\sigma$}{l3}
\fmfv{l.d=3,label=$\nu$}{l2}
\fmfv{l.d=3,label=$\mu$}{l1}
\fmfiv{d.s=circle,d.f=full,d.si=6}{point length(p1)/4 of p1}
\fmfiv{d.s=circle,d.f=full,d.si=6}{point length(p2)/4 of p2}
\fmfiv{d.s=circle,d.f=full,d.si=6}{point 0*length(p2)/4 of p2}
\fmfi{plain}{vpath (__phi,__r2) shifted (thick*(0,1.))}
\fmfi{plain}{vpath (__phi,__r2) shifted (thick*(0,-1.))}
\end{fmfgraph*}} \\[13mm]
\quad
&+\ \  \sum_{\underset{(\mu, \nu, \sigma)}{perm}}\quad
\parbox{50mm}{
\begin{fmfgraph*}(50,20)
\fmfrightn{r}{3}\fmfleftn{l}{3}
\fmf{phantom,tension=1.5}{l2,v4}
\fmf{phantom,tension=1}{v4,v2}
\fmf{phantom,tension=1}{v2,phi}
\fmf{phantom,tension=2.4}{phi,r2}
\fmffreeze
\fmf{phantom,tension=25}{v3,v2,v1}
\fmf{phantom,tension=1}{l2,v3}
\fmf{phantom,tension=1}{v1,r2}
\fmffreeze
\fmf{fermion,tension=.4,right,l.s=right,label=$p,,a$}{phi,v1}
\fmf{fermion,tension=.4,right,l.s=right,label=$p',,b$}{v1,phi}
\fmf{fermion,tension=.4,right=1,label=$k,,c$}{v3,v4}
\fmf{fermion,tension=.4,right=1,label=$k',,d$}{v4,v3}
\fmf{photon,left=.2,rubout=4}{l2,v3}
\fmfv{l.d=4,label=$\sigma$}{l3}
\fmfv{l.d=4,label=$\nu$}{l2}
\fmfv{l.d=4,label=$\mu$}{l1}
\fmffreeze
\fmf{phantom,left,tag=1}{v4,v3}
\fmf{phantom,right,tag=2}{v4,v3}
\fmfposition
\fmfipath{p[]}
\fmfiset{p1}{vpath1(__v4,__v3)}
\fmfiset{p2}{vpath2(__v4,__v3)}
\fmfi{photon}{point 0.27*length(p1) of p1 -- vloc(__l3)}
\fmfi{photon}{point 0.27*length(p2) of p2 -- vloc(__l1)}
\fmfiv{d.s=circle,d.f=full,d.si=6}{point 0.27*length(p1) of p1}
\fmfiv{d.s=circle,d.f=full,d.si=6}{point 0.27*length(p2) of p2}
\fmfiv{d.s=circle,d.f=full,d.si=6}{point .15*length(p2)/4 of p2}
\fmfiv{d.s=circle,d.f=full,d.si=6}{point 3*length(p2)/4 of p2}
\fmfiv{d.s=circle,d.f=full,d.si=6}{point 3*length(p1)/4 of p1}
\fmfv{d.sh=circle,d.f=empty,d.si=18,l=$\hspace{-3.2mm}\Phi$}{phi}
\fmfi{plain}{vpath (__phi,__r2) shifted (thick*(0,1.))}
\fmfi{plain}{vpath (__phi,__r2) shifted (thick*(0,-1.))}
\fmfv{d.s=square,d.f=full,d.si=8}{v2}
\end{fmfgraph*}}\
+\sum_{\underset{(\mu, \nu, \sigma)}{cyclic}}\quad
\parbox{50mm}{
\begin{fmfgraph*}(50,20)
\fmfrightn{r}{3}\fmfleftn{l}{3}
\fmf{phantom,tension=1.5}{l2,v4}
\fmf{phantom,tension=1}{v4,v2}
\fmf{phantom,tension=1}{v2,phi}
\fmf{phantom,tension=2.4}{phi,r2}
\fmffreeze
\fmf{phantom,tension=25}{v3,v2,v1}
\fmf{phantom,tension=1}{l2,v3}
\fmf{phantom,tension=1}{v1,r2}
\fmffreeze
\fmf{fermion,tension=.4,right,l.s=right,label=$p,,a$}{phi,v1}
\fmf{fermion,tension=.4,right,l.s=right,label=$\hspace{1cm}p',,b$}{v1,phi}
\fmf{fermion,tension=.4,right=1,label=$k,,c$}{v3,v4}
\fmf{fermion,tension=.4,right=1,label=$k',,d$}{v4,v3}
\fmfv{d.s=circle,d.f=full,d.si=6}{v4}
\fmf{photon,rubout=4}{l3,v3}
\fmf{photon}{l1,v4}
\fmf{photon}{l2,v4}
\fmfv{l.d=4,label=$\sigma$}{l3}
\fmfv{l.d=4,label=$\nu$}{l2}
\fmfv{l.d=4,label=$\mu$}{l1}
\fmffreeze
\fmf{phantom,right,tag=1}{v3,v4}
\fmf{phantom,left,tag=2}{v3,v4}
\fmfposition
\fmfipath{p[]}
\fmfiset{p1}{vpath1(__v3,__v4)}
\fmfiset{p2}{vpath2(__v3,__v4)}
\fmfiv{d.s=circle,d.f=full,d.si=6}{point length(p1)/4 of p1}
\fmfiv{d.s=circle,d.f=full,d.si=6}{point 3*length(p2)/4 of p2}
\fmfv{d.sh=circle,d.f=empty,d.si=18,l=$\hspace{-3.2mm}\Phi$}{phi}
\fmfi{plain}{vpath (__phi,__r2) shifted (thick*(0,1.))}
\fmfi{plain}{vpath (__phi,__r2) shifted (thick*(0,-1.))}
\fmfv{d.sh=circle,d.f=empty,d.si=18,l=$\hspace{-3.2mm}\Phi$}{phi}
\fmfi{plain}{vpath (__phi,__r2) shifted (thick*(0,1.))}
\fmfi{plain}{vpath (__phi,__r2) shifted (thick*(0,-1.))}
\fmfv{d.s=square,d.f=full,d.si=8}{v2}
\end{fmfgraph*}} \\[13mm]
&
+\ \  \sum_{\underset{(\mu, \nu, \sigma)}{cyclic}}\quad\quad
\parbox{50mm}{
\begin{fmfgraph*}(50,20)
\fmfrightn{r}{3}\fmfleftn{l}{3}
\fmf{phantom,tension=1.5}{l2,v4}
\fmf{phantom,tension=1}{v4,v2}
\fmf{phantom,tension=1}{v2,phi}
\fmf{phantom,tension=2.4}{phi,r2}
\fmffreeze
\fmf{phantom,tension=25}{v3,v2,v1}
\fmf{phantom,tension=1}{l2,v3}
\fmf{phantom,tension=1}{v1,r2}
\fmffreeze
\fmf{fermion,tension=.4,right,l.s=right,label=$p,,a$}{phi,v1}
\fmf{fermion,tension=.4,right,l.s=right,label=$p',,b$}{v1,phi}
\fmf{fermion,tension=.4,right=1,label=$k,,c$}{v3,v4}
\fmf{fermion,tension=.4,right=1,label=$k,,d$}{v4,v3}
\fmfv{d.s=circle,d.f=full,d.si=6}{v4}
\fmf{photon,rubout=4}{v3,l1}
\fmf{photon,rubout=4}{v2,l3}
\fmf{photon}{v4,l2}
\fmfv{l.d=4,label=$\sigma$}{l3}
\fmfv{l.d=4,label=$\nu$}{l1}
\fmfv{l.d=4,label=$\mu$}{l2}
\fmffreeze
\fmf{phantom,right,tag=1}{v3,v4}
\fmf{phantom,left,tag=2}{v3,v4}
\fmfposition
\fmfipath{p[]}
\fmfiset{p1}{vpath1(__v3,__v4)}
\fmfiset{p2}{vpath2(__v3,__v4)}
\fmfiv{d.s=circle,d.f=full,d.si=6}{point length(p1)/4 of p1}
\fmfiv{d.s=circle,d.f=full,d.si=6}{point length(p2)/4 of p2}
\fmfv{d.sh=circle,d.f=empty,d.si=18,l=$\hspace{-3.2mm}\Phi$}{phi}
\fmfi{plain}{vpath (__phi,__r2) shifted (thick*(0,1.))}
\fmfi{plain}{vpath (__phi,__r2) shifted (thick*(0,-1.))}
\fmfv{d.sh=circle,d.f=empty,d.si=18,l=$\hspace{-3.2mm}\Phi$}{phi}
\fmfi{plain}{vpath (__phi,__r2) shifted (thick*(0,1.))}
\fmfi{plain}{vpath (__phi,__r2) shifted (thick*(0,-1.))}
\fmfv{d.s=square,d.f=full,d.si=8}{v2}
\end{fmfgraph*}}\quad
+\quad\ \
\parbox{50mm}{
\begin{fmfgraph*}(50,20)
\fmfrightn{r}{3}\fmfleftn{l}{3}
\fmf{phantom,tension=1.5}{l2,v4}
\fmf{phantom,tension=1}{v4,v2}
\fmf{phantom,tension=1}{v2,phi}
\fmf{phantom,tension=2.4}{phi,r2}
\fmffreeze
\fmf{phantom,tension=25}{v3,v2,v1}
\fmf{phantom,tension=1}{l2,v3}
\fmf{phantom,tension=1}{v1,r2}
\fmffreeze
\fmf{fermion,tension=.4,right,l.s=right,label=$p,,a$}{phi,v1}
\fmf{fermion,tension=.4,right,l.s=right,label=$\hspace{1cm}p',,b$}{v1,phi}
\fmf{fermion,tension=.4,right=1,label=$k,,c$}{v3,v4}
\fmf{fermion,tension=.4,right=1,label=$k,,d$}{v4,v3}
\fmf{photon,rubout=4}{v1,l1}
\fmf{photon,rubout=4}{v1,l3}
\fmf{photon,rubout=4}{v	1,l2}
\fmfv{l.d=4,label=$\sigma$}{l3}
\fmfv{l.d=4,label=$\nu$}{l2}
\fmfv{l.d=4,label=$\mu$}{l1}
\fmffreeze
\fmf{phantom,right,tag=1}{v3,v4}
\fmf{phantom,left,tag=2}{v3,v4}
\fmfposition
\fmfipath{p[]}
\fmfiset{p1}{vpath1(__v3,__v4)}
\fmfiset{p2}{vpath2(__v3,__v4)}
\fmfiv{d.s=circle,d.f=full,d.si=6}{point length(p1)/4 of p1}
\fmfv{d.sh=circle,d.f=empty,d.si=18,l=$\hspace{-3.2mm}\Phi$}{phi}
\fmfi{plain}{vpath (__phi,__r2) shifted (thick*(0,1.))}
\fmfi{plain}{vpath (__phi,__r2) shifted (thick*(0,-1.))}
\fmfv{d.sh=circle,d.f=empty,d.si=18,l=$\hspace{-3.2mm}\Phi$}{phi}
\fmfi{plain}{vpath (__phi,__r2) shifted (thick*(0,1.))}
\fmfi{plain}{vpath (__phi,__r2) shifted (thick*(0,-1.))}
\fmfv{d.s=square,d.f=full,d.si=8}{v2}
\end{fmfgraph*}}
\end{align*}
\end{fmffile} 
\vspace{0mm}

\caption{\fign{fig:Amns} Illustration of the three-photon decay amplitude $A^{\mu\nu\sigma}$ as given by \eq{Amns-symb}. A black circled vertex where one wiggly $\mu$-labeled line (EM current) attaches to a solid line (quark) represents the
dressed quark EM vertex function $\Gamma^\mu$. Similarly, a black circled vertex where two (appropriately labeled) wiggly lines attach to a solid line represents the vertex function  $\Gamma^{\mu\nu}$. Black squared vertices with one, two, or three (appropriately labeled) attached wiggly lines represent the one, two, or three-times gauged BS kernels $K^\mu$, $K^{\mu\nu}$, or $K^{\mu\nu\sigma}$, respectively.}
\end{figure}

As for the one- and two-photon decay amplitudes $A^\mu$ and $A^{\mu\nu}$, the gauge invariance of the three-photon decay amplitude $A^{\mu\nu\sigma}$  rests crucially on the use of one universal BS kernel $K$ to calculate all the quantities required for its construction..
Thus, the same $K$ is to be used in constructing the bound state wave function $\Phi$ with \eq{BS}, the dressed quark propagator with \eq{DS}, and the dressed quark EM vertex $\Gamma^\mu$ with \eq{Gammamu-fsi}. The same is true for the vertex function $\Gamma^{\mu\nu}$ whose $K$-dependence is given by (see Appendix C for the derivation)
\begin{align}
\Gamma^{\mu\nu} &=\left( \Gamma^\mu S\Gamma^\nu + \Gamma^\nu S\Gamma^\mu\right) G_0 T +
\left(S^\mu K^\nu+S^\nu K^\mu+S K^{\mu\nu} \right) (1+G_0 T).   \eqn{Gammamunu}
\end{align}
The important feature of this expression is the appearance of the $q\q$ scattering $t$ matrix $T$ whose $K$ dependence is given by the BS equation, \eq{BST}. This $T$ dependence of $\Gamma^{\mu\nu}$ is not illustrated in \fig{fig:Amns} due to lack of space, yet it is of central importance to the gauge invariance of the three-photon decay amplitude (through the use of a universal $K$).
To illustrate this $T$ dependence more clearly, it is sufficient to consider the case where the kernel does not support charge flow, so that $K^\mu=K^\nu=K^\sigma=0$. 
In this case the three-photon decay amplitude becomes
\begin{align}
A^{\mu\nu\sigma} &=\sum_{perm}\mbox{Tr}\,  \Gamma^{\mu} S\Gamma^\nu S \Gamma^\sigma \Phi\nn
&+\sum_{perm}\mbox{Tr}\left[\left(\Gamma^\mu S\Gamma^\nu G_0 T\right) S\Gamma^\sigma 
+ \Gamma^\mu S \left(\Gamma^\nu S\Gamma^\sigma G_0 T\right)\right]\Phi  \eqn{Amns2}
\end{align}
which is illustrated in \fig{fig:Amns1}. It is interesting to compare the two-photon decay amplitude for the same case of no charge flow in the kernel  ($K^\mu=K^\nu=0$):
\be
A^{\mu\nu}=\sum_{perm}\mbox{Tr}\, \Gamma^\mu S \Gamma^\nu  \Phi   \eqn{Amn22}
\ee
where in this case the permutation sum is just over $\mu$ and $\nu$.
Thus the first term of \eq{Amns2} looks like a  direct analogue of the two-photon decay amplitude of \eq{Amn22}, in that both correspond to a quark loop diagram with attached photons. It is an important result of this paper that for 3-photon decay, gauge invariance requires the additional contribution of the second  term in \eq{Amns2} which corresponds to $q\q$ scattering after one photon is emitted by one of the constituents of the meson $\Phi$. It is noteworthy that our result corresponds to the one obtained in Refs.\ \cite{Goecke:2012qm}  and \cite{Dorokhov:2008pw} for light-by-light scattering, if we simply replace our $q\q$ bound state leg in \fig{fig:Amns1} by a photon. In this respect,
it is instructive to compare how  gauge invariance is achieved for light-by-light scattering in Refs.\ \cite{Goecke:2012qm} and \cite{Dorokhov:2008pw}. As in the present paper, the DSE and the gauging of equations technique were used in Ref.\ \cite{Goecke:2012qm}. By contrast, Dorokhov {\em et.\ al.} \cite{Dorokhov:2008pw,Dorokhov:2012qa,Dorokhov:2014iva} used an NJL-like model to derive a result, analogous to that of \fig{fig:Amns1}, by summing {\em all} leading-order terms of the $1/N_c$ expansion of the light-by-light scattering amplitude. In the latter case, gauge invariance is a consequence of the completeness of the LO contributions, even though the expression for the light-by-light scattering amplitude involves all orders of the strong interaction coupling constant. One can notice that the meson line in Fig.\ 3 of Ref.\ \cite{Dorokhov:2008pw}, which mimics the $q\bar q$ scattering  amplitude of our \fig{fig:Amns1}, represents the result of iteration of the kernel in the horizontal direction. In our case, this direction is dictated by the gauging procedure leading to the gauge invariant result. 
In the case of Ref.\ \cite{Dorokhov:2008pw}, iteration in the horizontal direction (resulting in a factor $N^2_c$ due to two traces over colors) is included because it is of LO in the $1/N_c$ expansion, while the one in the vertical direction (involving a factor $N_c$ due to only one trace over colors) is excluded because it is of NLO.
Evidently, the gauging of equations method has more possibilities as it allows one to start with any model for the dressed quark propagator, including ones in the framework of the NJL Lagrangian where the dressed quark propagator would involve different powers of $1/N_c$ (in addition to the strong interaction coupling constant). The simplest example would be the model where the quark self-energy $\Sigma$ is represented by the sum of diagrams consisting of the 
LO quark loop tadpole [of $(1/N_c)^0$ order], and the NLO quark-antiquark (meson) loop (of $1/N_c$ order).

\begin{figure}[t]
\begin{fmffile}{Amns1}
\begin{align*}
A^{\mu\nu\sigma}\ \ &=\ \  \sum_{\underset{(\mu, \nu, \sigma)}{perm}}\quad\quad
\parbox{35mm}{
\begin{fmfgraph*}(35,20)
\fmfleftn{l}{3}\fmfrightn{r}{3}
\fmf{phantom,tension=1}{l2,v}
\fmf{plain,tension=.4,right}{phi,v}
\fmf{plain,tension=.4,right}{v,phi}
\fmf{phantom,tension=1.8}{phi,r2}
\fmffreeze
\fmfv{d.sh=circle,d.f=empty,d.si=18,l=$\hspace{-3.2mm}\Phi$}{phi}
\fmf{phantom,left,tag=1}{v,phi}
\fmf{phantom,right,tag=2}{v,phi}
\fmfposition
\fmfipath{p[]}
\fmfiset{p1}{vpath1(__v,__phi)}
\fmfiset{p2}{vpath2(__v,__phi)}
\fmfi{photon}{point 0.43*length(p1) of p1 -- vloc(__l3)}
\fmfv{l.d=3,label=$\sigma$}{l3}
\fmfv{l.d=3,label=$\nu$}{l2}
\fmfv{l.d=3,label=$\mu$}{l1}
\fmfi{photon}{point 0.43*length(p2) of p2 -- vloc(__l1)}
\fmf{photon}{v,l2}
\fmfiv{d.s=circle,d.f=full,d.si=6}{point 0.43*length(p1) of p1}
\fmfiv{d.s=circle,d.f=full,d.si=6}{point 0.43*length(p2) of p2}
\fmfiv{d.s=circle,d.f=full,d.si=6}{point 0.43*length(p1)/2 of p1}
\fmfiv{d.s=circle,d.f=full,d.si=6}{point 0.43*length(p2)/2 of p2}
\fmfv{d.s=circle,d.f=full,d.si=6}{v}
\fmfi{plain,l.s=right,label=$$}{subpath (length(p1),length(p1)/2) of p1}
\fmfi{plain,l.s=right,label=$$}{subpath (length(p2)/2,length(p2)) of p2}
\fmfi{plain}{vpath (__phi,__r2) shifted (thick*(0,1.))}
\fmfi{plain}{vpath (__phi,__r2) shifted (thick*(0,-1.))}
\end{fmfgraph*}}\\[16mm]
\quad
&+\ \  \sum_{\underset{(\mu, \nu, \sigma)}{perm}}\quad
\parbox{50mm}{
\begin{fmfgraph*}(50,20)
\fmfrightn{r}{3}\fmfleftn{l}{3}
\fmf{phantom,tension=1.5}{l2,v4}
\fmf{phantom,tension=1}{v4,v2}
\fmf{phantom,tension=1}{v2,phi}
\fmf{phantom,tension=2.4}{phi,r2}
\fmffreeze
\fmf{phantom,tension=25}{v3,v2,v1}
\fmf{phantom,tension=1}{l2,v3}
\fmf{phantom,tension=1}{v1,r2}
\fmffreeze
\fmf{plain,tension=.4,right,l.s=right}{phi,v1}
\fmf{plain,tension=.4,right,l.s=right,label=$$}{v1,phi}
\fmf{plain,tension=.4,right=1,label=$$}{v3,v4}
\fmf{plain,tension=.4,right=1,label=$$}{v4,v3}
\fmfv{l.d=4,label=$\sigma$}{l3}
\fmfv{l.d=4,label=$\nu$}{l2}
\fmfv{l.d=4,label=$\mu$}{l1}
\fmffreeze
\fmf{phantom,left,tag=1}{v4,v3}
\fmf{phantom,right,tag=2}{v4,v3}
\fmf{phantom,right,tag=3}{phi,v1}
\fmf{phantom,tag=4}{l1,l3}
\fmfposition
\fmfipath{pp[]}
\fmfiset{pp1}{vpath1(__v4,__v3)}
\fmfiset{pp2}{vpath2(__v4,__v3)}
\fmfiset{pp3}{vpath3(__phi,__v1)}
\fmfi{photon}{point 0.30*length(pp1) of pp1 -- vloc(__l2)}
\fmfi{photon}{point 0.30*length(pp2) of pp2 -- vloc(__l1)}
\fmfi{plain,l.s=right,label=$$}{subpath (0,0.5*length(pp3)) of pp3}
\fmfi{photon}{%
 begingroup;
  clearxy; save p, t;
  path p[]; numeric t[];
  p3=vpath3(__phi,__v1); p4=vpath4(__l1,__l3);
  t1 = 0.46*length(p3);  z1 = point t1 of p3;
  t2 = 1*length(p4);  z2 = point t2 of p4;
   z1{direction t1 of p3 rotated -50}
 .. z2{direction t2 of p4 rotated 120}
endgroup
}
\fmfiv{d.s=circle,d.f=full,d.si=6}{point 0.70*length(pp3) of pp3}
\fmfiv{d.s=circle,d.f=full,d.si=6}{point 0.46*length(pp3) of pp3}
\fmfiv{d.s=circle,d.f=full,d.si=6}{point 0.30*length(pp1) of pp1}
\fmfiv{d.s=circle,d.f=full,d.si=6}{point 0.30*length(pp2) of pp2}
\fmfiv{d.s=circle,d.f=full,d.si=6}{point 5*length(pp1)/8 of pp1}
\fmfiv{d.s=circle,d.f=full,d.si=6}{point 5*length(pp2)/8 of pp2}
\fmfiv{d.s=circle,d.f=full,d.si=6}{point 0*length(pp2)/8 of pp2}
\fmfv{d.sh=circle,d.f=empty,d.si=18,l=$\hspace{-3.2mm}\Phi$}{phi}
\fmfi{plain}{vpath (__phi,__r2) shifted (thick*(0,1.))}
\fmfi{plain}{vpath (__phi,__r2) shifted (thick*(0,-1.))}
\fmfv{d.sh=circle,d.f=empty,d.si=18,l=$\hspace{-3.2mm}T$}{v2}
\end{fmfgraph*}}
\quad
+\ \  \sum_{\underset{(\mu, \nu, \sigma)}{perm}}\quad
\parbox{50mm}{
\begin{fmfgraph*}(50,20)
\fmfrightn{r}{3}\fmfleftn{l}{3}
\fmf{phantom,tension=1.5}{l2,v4}
\fmf{phantom,tension=1}{v4,v2}
\fmf{phantom,tension=1}{v2,phi}
\fmf{phantom,tension=2.4}{phi,r2}
\fmffreeze
\fmf{phantom,tension=25}{v3,v2,v1}
\fmf{phantom,tension=1}{l2,v3}
\fmf{phantom,tension=1}{v1,r2}
\fmffreeze
\fmf{plain,tension=.4,right,l.s=right}{phi,v1}
\fmf{plain,tension=.4,right,l.s=right,label=$$}{v1,phi}
\fmf{plain,tension=.4,right=1,label=$$}{v3,v4}
\fmf{plain,tension=.4,right=1,label=$$}{v4,v3}
\fmfv{l.d=4,label=$\sigma$}{l3}
\fmfv{l.d=4,label=$\nu$}{l2}
\fmfv{l.d=4,label=$\mu$}{l1}
\fmffreeze
\fmf{phantom,left,tag=1}{v4,v3}
\fmf{phantom,right,tag=2}{v4,v3}
\fmf{phantom,left,tag=3}{phi,v1}
\fmf{phantom,tag=4}{l1,l3}
\fmfposition
\fmfipath{pp[]}
\fmfiset{pp1}{vpath1(__v4,__v3)}
\fmfiset{pp2}{vpath2(__v4,__v3)}
\fmfiset{pp3}{vpath3(__phi,__v1)}
\fmfi{photon}{point 0.3*length(pp1) of pp1 -- vloc(__l3)}
\fmfi{photon}{point 0.26*length(pp2) of pp2 -- vloc(__l2)}
\fmfi{plain,l.s=right,label=$$}{subpath (0.5*length(pp3),0) of pp3}
\fmfi{photon}{ %
 begingroup;
  clearxy; save p, t;
  path p[]; numeric t[];
  p3=vpath3(__phi,__v1); p4=vpath4(__l1,__l3);
  t1 = 0.45*length(p3);  z1 = point t1 of p3;
  t2 = 0*length(p4);  z2 = point t2 of p4;
  z1{direction t1 of p3 rotated 50}
 .. z2{direction t2 of p4 rotated 60}
endgroup
}
\fmfiv{d.s=circle,d.f=full,d.si=6}{point 0.70*length(pp3) of pp3}
\fmfiv{d.s=circle,d.f=full,d.si=6}{point 0.46*length(pp3) of pp3}
\fmfiv{d.s=circle,d.f=full,d.si=6}{point 0.30*length(pp1) of pp1}
\fmfiv{d.s=circle,d.f=full,d.si=6}{point 0.30*length(pp2) of pp2}
\fmfiv{d.s=circle,d.f=full,d.si=6}{point 5*length(pp1)/8 of pp1}
\fmfiv{d.s=circle,d.f=full,d.si=6}{point 5*length(pp2)/8 of pp2}
\fmfiv{d.s=circle,d.f=full,d.si=6}{point 0*length(pp2)/8 of pp2}
\fmfv{d.sh=circle,d.f=empty,d.si=18,l=$\hspace{-3.2mm}\Phi$}{phi}
\fmfi{plain}{vpath (__phi,__r2) shifted (thick*(0,1.))}
\fmfi{plain}{vpath (__phi,__r2) shifted (thick*(0,-1.))}
\fmfv{d.sh=circle,d.f=empty,d.si=18,l=$\hspace{-3.2mm}T$}{v2}
\end{fmfgraph*}}
\end{align*}
\end{fmffile} 
\vspace{5mm}

\caption{\fign{fig:Amns1} Illustration of the three-photon decay amplitude $A^{\mu\nu\sigma}$ for the case $K^\mu=K^\nu=K^\sigma=0$, expressed in terms of fundamental $K$-dependent quantities: the dressed quark vertex function $\Gamma^\mu$ (black circled vertices), the $q\q$ BS $t$ matrix $T$, the bound state wave function $\Phi$, and connecting dressed quark propagators (solid lines with black circle decoration). Note that $t$ matrix T results from the iteration of the kernel $K$ in the horizontal direction in the drawing.
}
\end{figure}

The $3\gamma$ decay amplitude, \eq{Amns-symb}, with exposed indices and momentum variables has the form
\begin{align}\label{3gam-expos}
&A^{\mu\nu\sigma }=\sum_{perm}\int  \frac{d^4p}{(2\pi)^4}\mbox{Tr}\bigl[\Gamma^\mu (p+q_3+q_2)S(p+q_3+q_2)
\nn
&\hspace{3cm} \times \Gamma^\nu(p+q_3) S(p+q_3)\Gamma^\sigma (p)\Phi(P,p)\bigr]
\nn[2mm]
&+\sum_{cycle}\int \frac{d^4p}{(2\pi)^4}\mbox{Tr}\bigl\{\bigl[ \Gamma^{\mu\nu}(p+q_3) S(p+q_3) \Gamma^\sigma(p)
\nn
&\hspace{3cm}+\ \Gamma^\mu(p+q_2+q_3)  S(p+q_2+q_3) \Gamma^{\nu\sigma}(p)\bigr] \Phi(P,p)\bigr\}\nn[2mm]
&+\sum_{perm}\int \frac{d^4p\,d^4k}{(2\pi)^8} \sum_{abcdefgh}\bigl\{S^h_d(k+q_1+q_3)[\Gamma^\mu]^g_h(k+q_3) S^f_g(k+q_3)
[\Gamma^{\sigma}]^e_f(k)\nn
&\hspace{3cm}\times  S^c_e(k)
[K^\nu]^{da}_{cb}(k,k+q_1+q_3;p,p')\bigr\} \Phi_a^b(P,p)
\nn[2mm]
&+\sum_{cycle}\int \frac{d^4p\,d^4k}{(2\pi)^8} \sum_{abcdef}\bigl\{ S^f_d(k+q_1+q_2) [\Gamma^{\mu\nu}]^e_f(k) S^c_e(k)
[K^\sigma]^{da}_{cb}(k,k+q_1+q_2;p,p')
\nn
&\hspace{3cm}+S^f_d(k+q_1) [\Gamma^\mu]^e_f(k) S^c_e(k)
[K^{\nu\sigma}]^{da} _{cb}(k,k+q_1;p,p')
\bigr\} \Phi_a^b(P,p)
\nn[2mm]
&+\int \frac{d^4p\,d^4k}{(2\pi)^8} \sum_{abcd}S_d^c(k)
[K^{\mu\nu\sigma }]^{da}_{cb}(k,k;p,p') \Phi_a^b(P,p)
\end{align}
where $p+q_3+q_2+q_1=p'$.

The gauge invariance of \eq{3gam-expos} is expressed through the WT identity for the five-point Green function $S^{\mu\nu\sigma}$.
Analogously to \eq{Smunu} for the double-gauged quark propagator $S^{\mu\nu}$, the expression for $S^{\mu\nu\sigma}$ has the  form of \eq{3gam-expos} but with the bound state wave function $\Phi$ replaced by the two-body quark-antiquark Green function $G$. The WT identity for $S^{\mu\nu\sigma}$ reads
\be
q_3{}_\sigma S^{\mu\nu\sigma }(p)=i[eS^{\mu\nu}(p)-S^{\mu\nu}(p+q_3) e].   \eqn{Smns-WTI}
\ee
The transversality condition for the three-photon decay amplitude of \eq{3gam-expos} is then a consequence of \eq{Smns-WTI}:
\be
q_1{}_\mu A^{\mu\nu\sigma }=q_2{}_\nu A^{\mu\nu\sigma }=q_3{}_\sigma A^{\mu\nu\sigma }=0.
\ee
This current conservation relation and the analyticity of $A^{\mu\nu\sigma }$ in the vicinity of $q_i=0$ guarantees Low's theorem for soft photon emission.
Although  the WT identity (\ref{Smns-WTI}) is obvious by construction (because $S^{\mu\nu\sigma }$ is obtained by attaching  a photon to all places in $S^{\mu\nu}$) it can easily be derived analogously to the derivation given in Appendix \ref{WTI} for $S^{\mu\nu}$.

\section{Discussion}

In this paper we used the gauging of equations method to construct the three-photon decay amplitude of $q\q$  bound states modeled by the Dyson-Schwinger and  Bethe-Salpeter equations. 
An essential aspect of our formulation is that it applies to all possible models of the quark self-energy $\Sigma$, including those, that unlike QCD, have charge carriers other than dressed quarks.
 For the construction of photon decay amplitudes, this means using a BS equation for the 
dressed $\gamma q\q$ vertex function $\Gamma^\mu$, \eq{Gammamu-full}, whose inhomogeneous term consists
not only of the usual "bare" three-point vertex function $e\gamma^\mu$,  but also of term $S K^\mu$ involving the five-point vertex function $K^\mu$, as illustrated in \fig{fig:Fbar}. For gauge invariance, it is required that $K^\mu$ satisfy the WT identity (\ref{WTI-K}) relating it to the BS kernel $K$. This can be achieved either phenomenologically, or by explicitly gauging the field theoretic expression for $K$.

The final expression for the three-photon decay amplitude is given by \eq{Amns-symb} in terms of the input quantities consisting of the $q\q$ bound state wave function $\Phi$, the dressed quark propagator $S$, and the dressed quark EM three- and four-point vertex functions $\Gamma^\mu$ and $\Gamma^{\mu\nu}$, respectively. In turn, $\Gamma^{\mu\nu}$ is expressed in terms of $\Gamma^\mu$, the $q\q$ scattering amplitude $T$, and the gauged kernels $K^\mu$, $K^{\mu\nu}$ and $K^{\mu\nu\sigma}$, as in  \eq{Gammamunu}.
Equation (\ref{Amns-symb}) is illustrated in \fig{fig:Amns}. This expression for the three-photon decay amplitude is gauge invariant as long as all the above input quantities are determined [through \eqs{BS}, (\ref{DS}),  (\ref{Gammamu-fsi}),  and (\ref{BST}), respectively] using the one universal BS kernel $K$. An essential feature of our approach is that gauge invariance is achieved in the way prescribed by QFT; namely, through the attachment of currents to all possible places in all the Feynman diagrams contributing to the $q\q$ Green function.

Our approach for constructing the $3\gamma$ decay amplitude is general, and can be applied to  "currents" of any desired tensor structure, $1, \gamma_5, \gamma^\mu, \gamma^\mu{\bm \tau}, \gamma^\mu\gamma_5, \gamma^\mu\gamma^\nu$ etc.  Although for those currents that do not satisfy a continuity equation, $\partial^\mu J_{\mu\dots}=0$ (with a Noether-like definition of the $J_{0\dots}$ component), the issue of gauge invariance is irrelevant, the main idea of the method remains; namely, to maintain the general structure of the QFT by ensuring that a Green function $G^\mu$  is constructed by attaching a current line (labeled by $\mu$) to all possible places in every Feynman diagram contributing to $G$. Just this property of our theoretical approach should lead to a non-trivial quark-level description of matrix elements involving external photons, mesons, or other bosons associated with the above-mentioned "general" currents. 
Indeed, with sufficiently sophisticated models for the initial dressed quark propagator, our gauging procedure can produce very rich dynamics, including non-perturbative properties like unitarity, generation of intermediate few-body states, etc.  We have shown,\footnote{To be published.} for example, that it is possible to construct a model of the dressed quark propagator such that its triple-gauging by appropriate currents, leads to equations for a system of two quarks and two antiquarks, with unitary two-meson and diquark-antidiquark cuts, similar to our recently derived tetraquark equations \cite{Kvinikhidze:2014yqa}.

Thus, our approach is applicable not only to $3\gamma$ decay, but also to other processes involving one, two, three, or more bosons "attached" to quark propagators, as for example, in descriptions of electromagnetic form factors \cite{Biernat:2013aka}, 
Compton scattering \cite{Eichmann:2011ec} and the closely related electromagnetic and pionic corrections \cite{Kvinikhidze:1998tp, Kvinikhidze:2001xb}, three-meson decay \cite{Kamano:2011ih}, meson-meson scattering \cite{Heupel:2012ua, Kvinikhidze:2014yqa}, photon-photon scattering \cite{Goecke:2012qm}, etc.  Some further examples of such matrix elements have been investigated in Ref.\ \cite{Eichmann:2011ec}. As in the case of $3\gamma$ decay, the building blocks of the corresponding amplitudes for such processes are determined by a universal kernel given in the underlying DS and BS equations.

\begin{acknowledgments}
A.N.K is supported by the Georgian Shota Rustaveli National Science Foundation (grant
11/31). The work of Z.K.S. is supported by the Ministry of Education and Science of the Russian Federation and in part by Russian Federation President Grant for the support of scientific schools NSh-2479.2014.2 and by RFBR grant 13-02-00418-a.

\end{acknowledgments}

\appendix

\section{Conditions for gauge invariance}
\label{sec:consistency}

It is well recognised that matrix elements of the current operator, when taken between physical (on mass shell) states, need to satisfy current conservation. For the meson one-photon decay amplitude $A^\mu$ of \eq{Amudef}, for example, this requirement is expressed in momentum space as
\be
q_\mu A^\mu = q_\mu \la 0|J^\mu(0)|\Phi\ra=0.   \eqn{Amucc}
\ee
Gauge invariance, however, not only implies current conservation in the way just described, but it constitutes a much more restrictive condition. In the operator formalism of QFT, gauge invariance requires the equal-time commutation relation
\be
[\Psi(x),J^0(y)]|_{x_0=y_0} = \delta^3(\x-\y) \hat Q \Psi(x)
\ee
to be satisfied by the quantized fields [where $\hat Q$ is the charge operator associated with the field $\Psi(x)$], in addition to current conservation, $\partial_\mu J^\mu(x)=0$.  A practical example of this difference is the well known Siegert's theorem in quantum mechanics 
 which can be shown to follow from gauge invariance, but not from current conservation alone \cite{Naus:1996eb}.
In the current context, gauge invariance is expressed through the requirement that the corresponding  $(n+1)$-point Green function satisfy the WT identity, relating it to the $n$-point Green function. Indeed, current conservation, as expressed by \eq{Amucc}, can be achieved in many ways, including trivial ones; yet, only the way based on the WT identity  is physically meaningful, as it reveals the rich dynamics of the QFT related to the process in question.
The purpose of this appendix is to clarify the conditions under which a BS model  yields a three-point function $S^\mu$ that satisfies the WT identity, and thus a
photo-decay amplitude $A^\mu$ that is gauge invariant.

\subsection{Models of  \bm{$\Sigma$} where dressed quarks are the only charge carriers}
\label{QCD-model}

In the exact case of QCD where  $J^\mu(z)=\q(z) e\gamma^\mu q(z)$, the quark 3-point function $S^\mu$ is expressed in terms of the 4-point Green function $G$ as
 \begin{align}
 [S^\mu(y',y)]^i_j&=\la 0|Tq_j(y')\bar q^i(y)J^\mu(0)|0\ra\nn 
 &=\la0|Tq_j(y')\bar q^i(y)
\bar q^a(0)q_b(0)|0\ra[e \gamma^\mu]_a^b\nn
&=G^{ia}_{jb}(y',y;0,0) [e \gamma^\mu]_a^b , \eqn{Smu-4Green}
\end{align}
which in momentum space is
\begin{align}
 [S^\mu(p',p)]^i_j&=\int d^4y'\, d^4y\, e^{i(p'y'-p y)}\la 0|Tq_j(y')\bar q^i(y)J^\mu(0)|0\ra\nn  
 &=\int d^4y'\, d^4y\, d^4x'\, d^4x \frac{d^4 k'}{(2\pi)^4} \frac{d^4 k}{(2\pi)^4} e^{i(p'y'-p y-k'x'+k x )} \nn
 &\hspace{3cm}\times G^{ia}_{jb}(y',y;x',x) [e \gamma^\mu]_a^b \nn
&=\int  \frac{d^4 k}{(2\pi)^4} \,G^{ia}_{jb}(p',p;k',k) [e \gamma^\mu]_a^b .  \eqn{Smu-4Greenm}
\end{align}
Equation (\ref{Smu-4Greenm}) can be used to take residues at poles corresponding to $q\q$ bound states, in this way leading (upon conjugation) to \eq{Amu1} that relates $A^\mu$ to the bound state wave function $\Phi$. Being a direct consequence of exact QCD, it is evident that the $S^\mu$ of \eq{Smu-4Greenm} will satisfy the WT identity when the exact QCD expression for the 4-point Green function $G$ is used. On the other hand, the WT identity  for $S^\mu$ will generally not be satisfied when an arbitrary 4-point Green function $G$  is used in \eq{Smu-4Greenm}. It is easy to show, however, that gauge invariance is maintained if the following consistency condition between the quark dressing $\Sigma$ and the BS kernel $K$ is satisfied:\footnote{A similar condition  in the context of axial currents is discussed  in Ref.\  \cite{Fischer:2006ub}. Note that the $\Sigma$ and $K$ in question are the ones that determine the $G$ in \eq{Smu-4Greenm} via the BS equation.}
\be
 \Sigma(p')e -e \Sigma(p)=
\int  \frac{d^4k}{(2\pi)^4}\, K(p',p;k',k)[S(k')e -e S(k)]   . \eqn{consist}
\ee
\eq{consist} can be derived as follows. \eq{Smu-4Greenm} (with spinor indices omitted) implies
\begin{align}
 S^\mu(p',p)&=\int \frac{d^4k}{(2\pi)^4} G(p',p;k',k) \left[e\gamma^\mu\right] \nn
 &=
\int \frac{d^4k}{(2\pi)^4}  [G_0+G_0 KG](p',p;k',k) \left[e\gamma^\mu\right]
\nn
&=S(p') e \gamma^\mu S(p)+\int  \frac{d^4k}{(2\pi)^4} [G_0K](p',p;k',k)S^\mu(k',k) .  \eqn{Smuexact}
\end{align}
Comparing this with 
 \begin{align}
 S^\mu(p',p)&=S(p') \Gamma^\mu(p',p) S(p),\\[3mm]
 &=S(p')[ e \gamma^\mu+\Sigma^\mu(p',p)] S(p),
 \end{align}
one obtains
 \begin{align}
 \Sigma^\mu(p',p)& =
\int  \frac{d^4k}{(2\pi)^4} K(p',p;k',k)S^\mu (k',k) .  \eqn{cons2}
\end{align}
Equation (\ref{cons2}) is equivalent to the well known definition of the $q\q$ BS kernel in terms of the functional derivative of the self-energy $\Sigma$ with respect to the dressed quark propagator  $S$  \cite{Munczek:1994zz},\footnote{{The technique
given by \eq{dSig-dS} is often called {\it cutting} since graphically it corresponds to the cutting of a dressed quark line \cite{Heupel:2014ina}. As such, it is similar to the U-gauging procedure proposed for derivation of generalized parton distributions, which by contrast, corresponds to the cutting of bare quark lines \cite{Kvinikhidze:2004dy}.}}

\be
    K(x_1,x_2,y_1,y_2) = \frac{\delta \Sigma(x_1-x_2)}{\delta S(y_1-y_2)}. \eqn{dSig-dS}
\ee
Note how  \eq{cons2} makes explicit that in this approach, photons attach only to dressed quark propagators in $\Sigma$.
Then using the WT identities for $\Sigma^\mu$ and $S^\mu$, one obtains
\begin{align}
i q_\mu\Sigma^\mu(p',p)&=\Sigma(p') e - e \Sigma(p)\nn
 &=
\int  \frac{d^4k}{(2\pi)^4} K(p',p;k',k)[S(k')e -e S(k)] .\eqn{cons3}
\end{align}
Assuming that the consistency condition of \eq{consist} is satisfied, one can now check the implication for the 3-point function $S^\mu$, when calculated using \eq{Smu-4Greenm} with $G$ modeled via $K$ and the BS equation. Indeed, \eq{Smu-4Greenm} (with spinor indices omitted) implies \eq{Smuexact},
which results in the following linear equation for $q_\mu S^\mu$:
\begin{align}
q_\mu S^\mu(p',p)&=S(p') i e \left[S_0^{-1}(p')-S_0^{-1}(p)\right]S(p)\nn
&+S(p')\left[\int  \frac{d^4k}{(2\pi)^4} K(p',p;k',k)q_\mu S^\mu(k',k)\right]S(p).
\end{align}
Given \eq{consist}, and the fact that $S^{-1}=S_0^{-1}-\Sigma$, the solution for $q_\mu S^\mu$ is
\be
q_\mu S^\mu(p',p) =i\left[e S(p)-S(p')e\right].
\ee
It is possible to use the consistency relation (\ref{consist}) in modeling the meson bound-state wave function and the gauge invariant 3 photon decay amplitude. In this case \eq{consist} would be considered as the equation for the quark propagator $S$ corresponding to a given $q\bar q$ interaction kernel $K$. It is easy to see, however, that such an equation is not solvable for an arbitrary kernel $K$, and thereby imposes a strong constraint on its form.  For example, the left hand side of \eq{consist} requires its right hand side (RHS) to be a difference of functions, $f(p')-f(p)$, which already is a strong constraint on $K$.

\begin{figure}[b]
\begin{fmffile}{DD}
\begin{align*} (a)&\quad\quad
\parbox{46mm}{
\begin{fmfgraph*}(46,20)
\fmfstraight
\fmfleftn{l}{2}\fmfrightn{r}{2}
\fmfset{curly_len}{2.5mm}
\fmf{fermion,tension=1.5,l.s=right,label=$k'$}{r2,u2}
\fmf{plain,tension=.7,l.s=right,label=$p'-k_2$}{u2,u1}
\fmf{fermion,tension=1.5,l.s=right,label=$p'$}{u1,l2}
\fmf{fermion,tension=1.5,l.s=right,label=$p$}{l1,v1}
\fmf{plain,tension=.7,l.s=right,label=$p-k_1$}{v1,v2}
\fmf{fermion,tension=1.5,l.s=right,label=$k$}{v2,r1}
\fmffreeze
\fmf{phantom}{u2,u,u1}
\fmf{phantom}{v2,v,v1}
\fmfv{d.s=circle,d.f=full,d.si=6}{u}
\fmfv{d.s=circle,d.f=full,d.si=6}{v}
\fmfv{l.a=90,label=$F^\sigma$}{u1}
\fmfv{l.a=90,label=$F^\alpha$}{u2}
\fmfv{l.a=-90,label=$F^\delta$}{v1}
\fmfv{l.a=-90,label=$F^\beta$}{v2}
\fmf{gluon}{v1,u2}
\fmf{gluon,rubout}{v2,u1}
\end{fmfgraph*}} 
\ +\
\parbox{46mm}{
\begin{fmfgraph*}(46,20)
\fmfstraight
\fmfleftn{l}{2}\fmfrightn{r}{2}
\fmfset{curly_len}{2.5mm}
\fmf{fermion,tension=1.5,l.s=right,label=$k'$}{r2,u2}
\fmf{plain,tension=1.2}{u2,uc,u1}
\fmf{fermion,tension=1.5,l.s=right,label=$p'$}{u1,l2}
\fmf{fermion,tension=1,l.s=right,label=$p$}{l1,vc}
\fmf{fermion,tension=1,l.s=right,label=$k$}{vc,r1}
\fmffreeze
\fmf{phantom}{u2,u,uc}
\fmf{phantom}{uc,v,u1}
\fmfv{d.s=circle,d.f=full,d.si=6}{u}
\fmfv{d.s=circle,d.f=full,d.si=6}{v}
\fmfv{l.a=-90,label=$F^\sigma$}{u1}
\fmfv{l.a=-90,label=$F^\beta$}{u2}
\fmfv{l.a=-90,label=$F^\delta$}{vc}
\fmfv{l.a=90,label=$F^\alpha$}{uc}
\fmf{gluon,l.d=10,label=$\uparrow k_1$}{vc,uc}
\fmf{gluon,right=.7,l.d=10,label=$\leftarrow k_2$}{u2,u1}
\end{fmfgraph*}} 
\ +\
\parbox{46mm}{
\begin{fmfgraph*}(46,20)
\fmfstraight
\fmfleftn{l}{2}\fmfrightn{r}{2}
\fmfset{curly_len}{2.5mm}
\fmf{fermion,tension=1,l.s=right,label=$k'$}{r2,uc}
\fmf{fermion,tension=1,l.s=right,label=$p'$}{uc,l2}
\fmf{fermion,tension=1.5,l.s=right,label=$p$}{l1,v1}
\fmf{plain,tension=1.2}{v1,vc,v2}
\fmf{fermion,tension=1.5,l.s=right,label=$k$}{v2,r1}
\fmffreeze
\fmf{phantom}{v2,u,vc}
\fmf{phantom}{vc,v,v1}
\fmfv{d.s=circle,d.f=full,d.si=6}{u}
\fmfv{d.s=circle,d.f=full,d.si=6}{v}
\fmfv{l.a=90,label=$F^\delta$}{v1}
\fmfv{l.a=90,label=$F^\alpha$}{v2}
\fmfv{l.a=-90,label=$F^\beta$}{vc}
\fmfv{l.a=90,label=$F^\sigma$}{uc}
\fmf{gluon,l.d=10,label=$\uparrow k_2$}{vc,uc}
\fmf{gluon,right=.7,l.d=10,label=$k_1 \rightarrow $}{v1,v2}
\end{fmfgraph*}} \\[23mm]
(b) &\quad\quad
\parbox{100mm}{
\begin{fmfgraph*}(100,20)
\fmfstraight
\fmfleftn{l}{1}\fmfrightn{r}{1}
\fmfset{curly_len}{2.5mm}
\fmf{fermion,tension=1,l.s=right,label=$p$}{r1,v4}
\fmf{plain,tension=1,l.s=right,label=$p-k_1$}{v4,v3}
\fmf{plain,tension=1,l.s=right,label=$$}{v3,v2}
\fmf{plain,tension=1,l.s=right,label=$p-k_2$}{v2,v1}
\fmf{fermion,tension=1,l.s=right,label=$p$}{v1,l1}
\fmffreeze
\fmf{phantom}{v1,a,v2}
\fmf{phantom}{v2,b,v3}
\fmf{phantom}{v3,c,v4}
\fmfv{d.s=circle,d.f=full,d.si=6}{a}
\fmfv{d.s=circle,d.f=full,d.si=6}{b}
\fmfv{d.s=circle,d.f=full,d.si=6}{c}
\fmfv{l.a=-90,label=$F^\sigma$}{v1}
\fmfv{l.a=-90,label=$F^\alpha$}{v2}
\fmfv{l.a=-90,label=$F^\beta$}{v3}
\fmfv{l.a=-90,label=$F^\delta$}{v4}
\fmf{gluon,right=.7,l.d=10,label=$\leftarrow k_2$}{v3,v1}
\fmf{gluon,right=.7,rubout,l.d=10,label=$\leftarrow k_1$}{v4,v2}
\end{fmfgraph*}} 
\end{align*}
\end{fmffile}
\vspace{-8mm}
\caption{\fign{DD} Illustration of (a) the two-gluon exchange BS kernel $K$ and (b) the two-gluon exchange quark self-energy $\Sigma$, that together satisfy the consistency condition, \eq{consist}. Here the quark-gluon vertices are bare but the quark propagators are dressed. }
\end{figure}
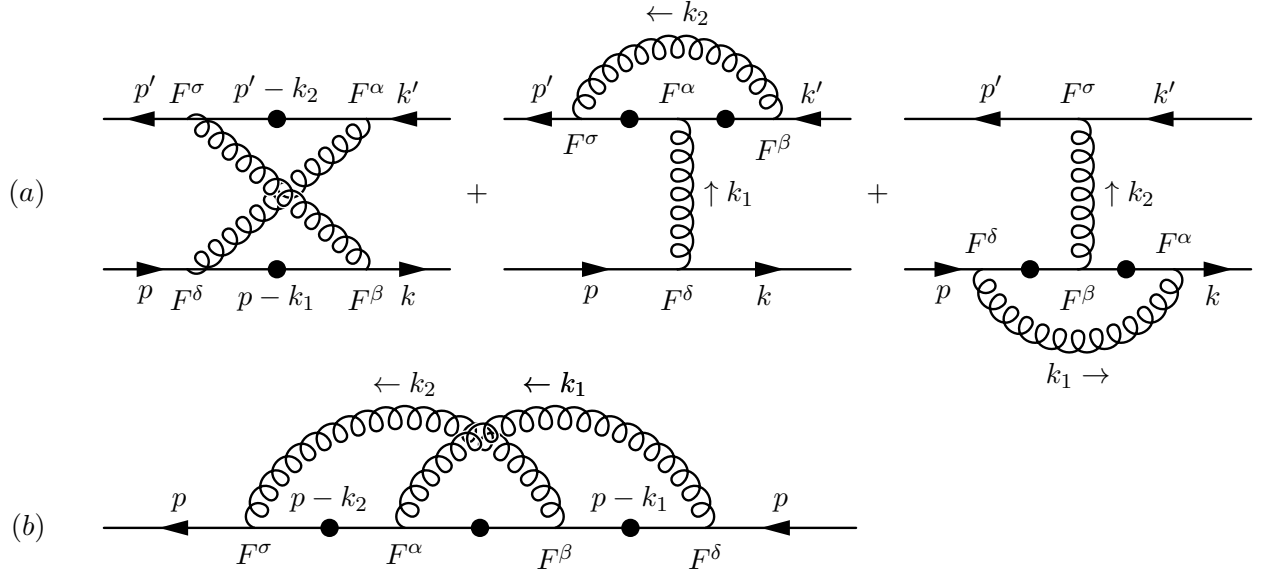

It is obvious that \eq{consist} is satisfied in the case of exact QCD; however, here we are interested to explore this condition for {\em models} where, as in exact QCD, the only charge-carrying constituents are dressed quarks. Thus, one needs to find appropriate model kernels for which \eq{consist} has a solution for $S$ (or corresponding $\Sigma$).  The most practical (and  at the same time interesting) such kernel is a local one with no charge exchange, $K(p',p;k',k)=D_{\alpha\beta}(p-k)F^\alpha \otimes F^\beta$, (e.g.,  an appropriately modified one-gluon exchange diagram with bare vertices, as in Ref.\ \cite{Maris:1999nt}) in which case the dressed quark propagator satisfies the  DS equation
 \be
 S^{-1}(p)= S_0^{-1}(p)-\int\frac{d^4k}{(2\pi)^4}D_{\alpha\beta}(p-k)F^\alpha S(k)F^\beta .
 \ee
It is instructive to check \eq{consist} in the more complicated case where $K$ is a two-gluon exchange BS kernel  with dressed internal quark propagators. In particular, \eq{consist} is satisfied for $K$ defined as the sum of the three diagrams illustrated in \fig{DD}(a).  None of theses diagrams, on their own, satisfy \eq{consist} - they do not even generate a difference $f(p')-f(p)$. Such a difference results from a cancellation in the sum of the three terms; indeed, this sum gives:
\begin{align}
\int &\frac{d^4k}{(2\pi)^4} K(p',p;k',k)[S(k')-S(k)]=
\int \frac{d^4k_1}{(2\pi)^4}\frac{d^4k_2}{(2\pi)^4}  D_{\delta\alpha}(k_1)D_{\beta\sigma}(k_2)
\nn
&\times F^\sigma\left\{ S(p'-k_2)
 F^\alpha [\underline{S(p'-k_1-k_2)}-\underline{\underline{S(p-k_1-k_2)}}] F^\beta S(p-k_1)\right.
 \nn
 &\hspace{5mm}+\left. S(p'-k_2)
 F^\alpha S(p'-k_1-k_2) F^\beta [S(p'-k_1)-\underline{S(p-k_1)}]\right.
 \nn
 &\hspace{5mm}+\left. [\underline{\underline{S(p'-k_2)}}-S(p-k_2)]
 F^\alpha S(p-k_1-k_2) F^\beta S(p-k_1)\right\}F^\sigma 
\nn[3mm]
=
\int& \frac{d^4k_1}{(2\pi)^4}\frac{d^4k_2}{(2\pi)^4}   D_{\delta\alpha}(k_1)D_{\beta\sigma}(k_2)
\nn
&\hspace{1mm}\times F^\sigma\left\{ S(p'-k_2)
 F^\alpha S(p'-k_1-k_2) F^\beta S(p'-k_1)\right.
 \nn
 &\hspace{1mm}-\left. S(p-k_2)
 F^\alpha S(p-k_1-k_2) F^\beta S(p-k_1)\right\}F^\sigma
=\Sigma(p')-\Sigma(p) \eqn{consist-2}
 \end{align}\newpage

\noindent
where $\Sigma(p)$ is illustrated in \fig{DD}(b).
In \eq{consist-2} the underlined  terms cancel each other, as do the double-underlined ones. The result is just the consistency relation (\ref{consist}) which guarantees the gauge invariance of the model defined by \fig{DD}.

To achieve this consistency condition for {\em any} model of $\Sigma$ where dressed quarks are the only charge carriers, we can use the gauging of equations approach. Namely, we start with such a model $\Sigma$, and then obtain the BS kernel (that generates the $q\bar q$ bound state) by gauging $\Sigma$. Once $\Sigma^\mu$ is determined, then \eq{cons2} can be used to identify $K$. This procedure is easily demonstrated on the example of the  $\Sigma$ of \fig{DD}(b), as follows:
\begin{align}
\Sigma(p) =\int &\frac{d^4k_1}{(2\pi)^4}\frac{d^4k_2}{(2\pi)^4} D_{\delta\alpha}(k_1)D_{\beta\sigma}(k_2)\nn[2mm]
 \times& F^\sigma  \left[ S(p-k_2)
 F^\alpha S(p-k_1-k_2) F^\beta S(p-k_1)\right] F^\sigma  .  \eqn{Sigma-2gamma}
\end{align}
Then gauging this equation gives
\begin{align}
&\Sigma^\mu(p) =\int \frac{d^4k}{(2\pi)^4} K(p',p;k',k)S^\mu(k) \nn[2mm]
&=
\int \frac{d^4k_1}{(2\pi)^4}\frac{d^4k_2}{(2\pi)^4} D_{\delta\alpha}(k_1)D_{\beta\sigma}(k_2)\nn
&\hspace{1cm} \times F^\sigma\left[ S(p'-k_2) F^\alpha S^\mu(p-k_1-k_2) F^\beta S(p-k_1)\right.
 \nn
 &\hspace{2cm}+\left. S(p'-k_2)
 F^\alpha S(p'-k_1-k_2) F^\beta S^\mu(p-k_1)\right.
 \nn
 &\hspace{2cm}+\left.S^\mu(p-k_2)
 F^\alpha S(p-k_1-k_2) F^\beta S(p-k_1)\right]F^\sigma .
\end{align}
Changing variables so that $S^\mu(k)$ appears in each term, we obtain
\begin{align*}
&\Sigma^\mu(p)=
\int \frac{d^4k}{(2\pi)^4}\biggl[ \int \frac{d^4k_1}{(2\pi)^4} D_{\delta\alpha}(k_1)D_{\beta\sigma}(p-k_1-k)  \nn
&\hspace{3cm} \times F^\sigma S(p'-p+k_1+k) F^\alpha \underline{S^\mu(k)} F^\beta S(p-k_1)F^\sigma
 \nn[1mm]
&\hspace{2cm} + \int \frac{d^4k_2}{(2\pi)^4} D_{\delta\alpha}(p-k)D_{\beta\sigma}(k_2) \nn
&\hspace{3cm} \times F^\sigma S(p'-k_2)
 F^\alpha S(p'-p+k-k_2) F^\beta \underline{S^\mu(k)}F^\sigma \nn[1mm]
 &\hspace{2cm}+\int  \frac{d^4k_1}{(2\pi)^4} D_{\delta\alpha}(k_1)D_{\beta\sigma}(p-k)\nn
 &\hspace{3cm}  \times F^\sigma \underline{S^\mu(k)}
 F^\alpha S(k-k_1) F^\beta S(p-k_1)F^\sigma \biggr]
 \end{align*}
 \begin{align}
 &  =
\int \frac{d^4k}{(2\pi)^4} \biggl[ \int \frac{d^4k_1}{(2\pi)^4} D_{\delta\alpha}(k_1)D_{\beta\sigma}(p-k_1-k)  \nn
& \hspace{3cm} \times F^\sigma S(p'-p+k_1+k) F^\alpha \otimes F^\sigma S(p-k_1) F^\beta 
 \nn[1mm]
&\hspace{2cm} + \int \frac{d^4k_2}{(2\pi)^4} D_{\delta\alpha}(p-k)D_{\beta\sigma}(k_2) \nn
&\hspace{3cm} \times F^\sigma S(p'-k_2)
 F^\alpha S(p'-p+k-k_2) F^\beta \otimes F^\sigma \nn[1mm]
 &\hspace{2cm}+\int  \frac{d^4k_1}{(2\pi)^4} D_{\delta\alpha}(k_1)D_{\beta\sigma}(p-k)\nn
 &\hspace{3cm}  \times F^\sigma  \otimes
 F^\sigma  S(p-k_1)  F^\beta S({k-k_1})    F^\alpha \biggr]  S^\mu(k).
 \eqn{K-gaug-Sig}
 \end{align}
The kernel $K$ corresponding to the model $\Sigma$ is thus defined by the expression within the square brackets in \eq{K-gaug-Sig}, and is seen to correspond exactly to the three diagrams of \fig{DD}(a). Once the model BS kernel has been determined in this way, one can use \eq{Amns2} to evaluate the three-photon decay amplitude. 

It is worth noting that \eq{consist}, together with axial WT identities, form the foundations
for constructing kernels that are used in contemporary hadron spectrum calculations;  for example, the model of \fig{DD} has already been solved in Ref.\ \cite{Williams:2009wx}.

\subsection{Models of \bm{$\Sigma$} where dressed quarks are not the only charge carriers}
\label{general-model}

Although \eq{Smu-4Greenm} is the correct expression for $S^\mu$ in exact QCD, there is 
no fundamental requirement, not even the requirement of gauge invariance, for retaining this expression when using models. As noted above, the use of \eq{Smu-4Greenm}  requires the quark self-energy $\Sigma$ to be a functional of just one charge-carrying object: the dressed quark propagator $S$. In such a case, external photons can only attach to the dressed quark propagators within $\Sigma$.
But one could have a reasonable model  of $\Sigma$ which depends also on other charge-carrying objects, and to which external photons can attach. In this case \eq{Smu-4Greenm} would no longer hold. Important cases of such models are provided by the QCD motivated effective field theories, of which  the NJL model \cite{Plant:1997jr,Plant:2000ty} is a notable example.

In order to demonstrate a simple model where photons do not attach just to dressed quark propagators, we consider a $\Sigma$ that is derived from \fig{DD}(b) by replacing the two outer internal dressed propagators by bare ones:
\begin{align}
\Sigma(p) =\int &\frac{d^4k_1}{(2\pi)^4}\frac{d^4k_2}{(2\pi)^4} D_{\delta\alpha}(k_1)D_{\beta\sigma}(k_2)\nn[2mm]
 \times& F^\sigma  \left[ S_0(p-k_2) F^\alpha S(p-k_1-k_2) F^\beta S_0(p-k_1)\right] F^\sigma \nn[3mm]
 \equiv\int&\frac{d^4k}{(2\pi)^4}V(p,k) S(k)  .  \eqn{Sigma-2gam-1}
\end{align}
\eq{Sigma-2gam-1} is derived from \eq{Sigma-2gamma} via the replacements  
$S(p-k_2)\rightarrow S_0(p-k_2)$ and $S(p-k_1)\rightarrow S_0(p-k_1)$, which turns $\Sigma$ into a linear functional of the dressed quark propagator~$S$. Gauging \eq{Sigma-2gam-1} one obtains
\begin{align}
&\Sigma^\mu(p) 
=
\int \frac{d^4k_1}{(2\pi)^4}\frac{d^4k_2}{(2\pi)^4} D_{\delta\alpha}(k_1)D_{\beta\sigma}(k_2)\nn
&\hspace{1cm} \times F^\sigma\bigl[ S_0(p'-k_2) F^\alpha S^\mu(p-k_1-k_2) F^\beta S_0(p-k_1)
 \nn
&\hspace{2cm}+ S_0(p'-k_2) F^\alpha S(p'-k_1-k_2) F^\beta S_0^\mu(p-k_1)
 \nn
&\hspace{2cm}+ S_0^\mu(p-k_2) F^\alpha S(p-k_1-k_2) F^\beta S_0(p-k_1)\bigr]F^\sigma 
\nn[2mm]
&=
\int \frac{d^4k_1}{(2\pi)^4}\frac{d^4k}{(2\pi)^4} \biggl\{  D_{\delta\alpha}(k_1)D_{\beta\sigma}(p-k_1-k)\nn
&\hspace{1cm} \times F^\sigma\bigl[ S_0(p'-p+k_1+k) F^\alpha S^\mu(k) F^\beta S_0(p-k_1)
 \nn
&\hspace{2cm}+ S_0^\mu(k+k_1) F^\alpha S(k) F^\beta S_0(p-k_1)\bigr]F^\sigma 
\nn[2mm]
&\hspace{1cm} + D_{\delta\alpha}(k_1)D_{\beta\sigma}(p'-k_1-k)\nn
&\hspace{1cm} \times F^\sigma\bigl[  S_0(k+k_1) F^\alpha S(k) F^\beta S_0^\mu(p-k_1)
\bigr]F^\sigma \biggr\}
\nn[2mm]
&=\int \frac{d^4k}{(2\pi)^4} K_X(p',p;k',k)S^\mu(k)+\int  \frac{d^4k}{(2\pi)^4} K_X^\mu(p',p;k,k)S(k)
\eqn{X-gaug-Sig}
\end{align}
where $K_X(p',p;k',k)$ is the crossed two-gluon exchange BS kernel, and $K^\mu_X(p',p;k',k)$ is the gauged crossed two gluon exchange BS kernel, so that they are related by the WT identity
\be
q_\mu K^\mu_X=[\hat{e}, K_X].
\ee
Equation (\ref{X-gaug-Sig}) is illustrated in \fig{DX}.
The dynamical origin of this WT identity  [in that $K^\mu_X(p',p;k',k)$ is obtained by attaching photons everywhere inside $K_X(p',p;k',k)$], is important . 
It is easy to check that
\be
V(p,k)=K_X(p,p;k,k).
\ee 
The quark-photon three-point function $S^\mu=S\left(e\gamma^\mu + \Sigma^\mu\right)S$ in this model can thus be expressed as
\begin{align}
S^\mu(p)&=S(p')\biggl[ F^\mu(p',p) +\int \frac{d^4k}{(2\pi)^4} K_X(p',p;k',k)S^\mu(k)\biggr]S(p)   \eqn{hello}
\end{align}
where
\be
F^\mu(p',p) =  e\gamma^\mu+\int \frac{d^4k}{(2\pi)^4} K_X^\mu(p',p;k,k)S(k).
\ee
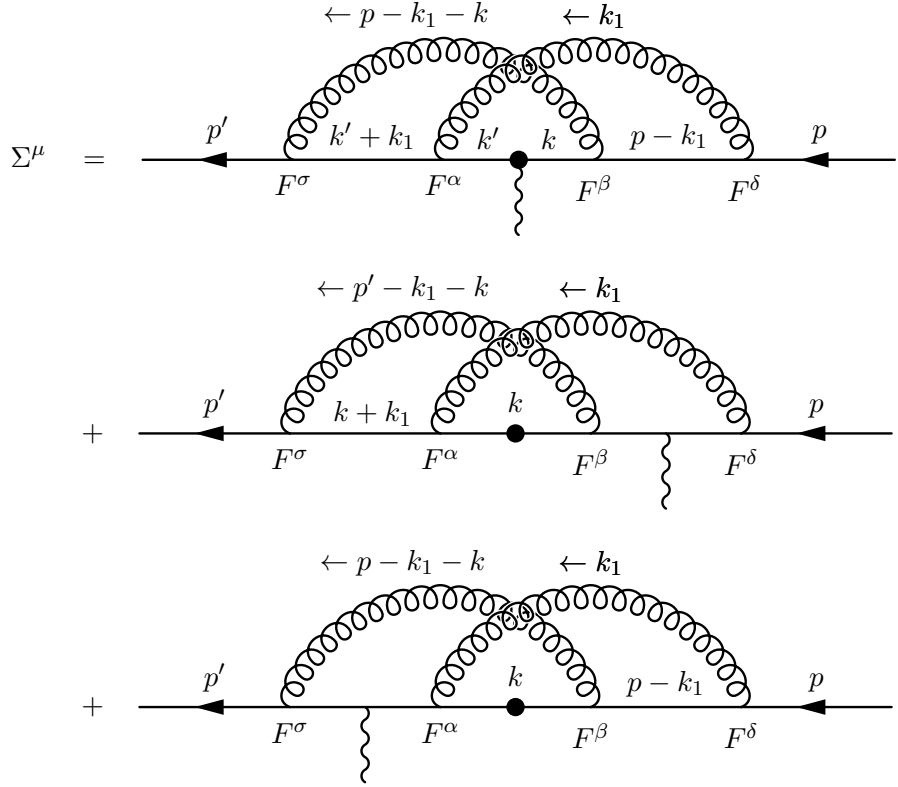
\begin{figure}[t]
\begin{fmffile}{DX}
\begin{align*} \Sigma^\mu \quad &=\quad
\parbox{100mm}{
\begin{fmfgraph*}(100,20)
\fmfstraight
\fmfleftn{l}{3}\fmfrightn{r}{3}\fmfbottomn{b}{3}
\fmfset{curly_len}{2.5mm}
\fmf{fermion,tension=1,l.s=right,label=$p$}{r2,v4}
\fmf{plain,tension=1,l.d=4,l.s=right,label=$p-k_1$}{v4,v3}
\fmf{plain,tension=1,l.s=right,label=$$}{v3,v2}
\fmf{plain,tension=1,l.d=4,l.s=right,label=$\hspace{1mm}Òk'+k_1$}{v2,v1}
\fmf{fermion,tension=1,l.s=right,label=$p'$}{v1,l2}
\fmffreeze
\fmf{phantom}{v1,a,v2}
\fmf{phantom}{v2,b,v3}
\fmf{phantom}{v3,c,v4}
\fmfv{d.s=circle,d.f=full,d.si=6}{b}
\fmfv{l.a=-90,label=$F^\sigma$}{v1}
\fmfv{l.a=-90,label=$F^\alpha$}{v2}
\fmfv{l.a=-90,label=$F^\beta$}{v3}
\fmfv{l.a=-90,label=$F^\delta$}{v4}
\fmffreeze
\fmf{phantom,l.d=4,label=$\hspace{-3mm}k$}{v3,b}
\fmf{phantom,l.d=4,label=$\hspace{3mm}k'$}{b,v2}
\fmf{gluon,right=.7,l.d=10,label=$\hspace{-1cm}\leftarrow p-k_1-k$}{v3,v1}
\fmf{gluon,right=.7,rubout,l.d=10,label=$\leftarrow k_1$}{v4,v2}
\fmf{photon,tension=0}{b,b2}
\end{fmfgraph*}} \\[15mm]
&+\quad
\parbox{100mm}{
\begin{fmfgraph*}(100,20)
\fmfstraight
\fmfleftn{l}{3}\fmfrightn{r}{3}\fmfbottomn{b}{11}
\fmfset{curly_len}{2.5mm}
\fmf{fermion,tension=1,l.s=right,label=$p$}{r2,v4}
\fmf{plain,tension=1,l.s=right,label=$$}{v4,v3}
\fmf{plain,tension=1,l.d=8,l.s=right,label=$k$}{v3,v2}
\fmf{plain,tension=1,l.d=3,l.s=right,label=$\hspace{2mm}k+k_1$}{v2,v1}
\fmf{fermion,tension=1,l.s=right,label=$p'$}{v1,l2}
\fmffreeze
\fmf{phantom}{v1,a,v2}
\fmf{phantom}{v2,b,v3}
\fmf{phantom}{v3,c,v4}
\fmffreeze
\fmfv{d.s=circle,d.f=full,d.si=6}{b}
\fmfv{l.a=-90,label=$F^\sigma$}{v1}
\fmfv{l.a=-90,label=$F^\alpha$}{v2}
\fmfv{l.a=-90,label=$F^\beta$}{v3}
\fmfv{l.a=-90,label=$F^\delta$}{v4}
\fmf{gluon,right=.7,l.d=10,label=$\hspace{-1cm}\leftarrow p'-k_1-k$}{v3,v1}
\fmf{gluon,right=.7,rubout,l.d=10,label=$\leftarrow k_1$}{v4,v2}
\fmf{photon,tension=0}{c,b8}
\end{fmfgraph*}} \\[15mm]
&+\quad
\parbox{100mm}{
\begin{fmfgraph*}(100,20)
\fmfstraight
\fmfleftn{l}{3}\fmfrightn{r}{3}\fmfbottomn{b}{11}
\fmfset{curly_len}{2.5mm}
\fmf{fermion,tension=1,l.s=right,label=$p$}{r2,v4}
\fmf{plain,tension=1,l.d=4,l.s=right,label=$p-k_1$}{v4,v3}
\fmf{plain,tension=1,l.d=8,l.s=right,label=$k$}{v3,v2}
\fmf{plain,tension=1,l.s=right,label=$$}{v2,v1}
\fmf{fermion,tension=1,l.s=right,label=$p'$}{v1,l2}
\fmffreeze
\fmf{phantom}{v1,a,v2}
\fmf{phantom}{v2,b,v3}
\fmf{phantom}{v3,c,v4}
\fmffreeze
\fmfv{d.s=circle,d.f=full,d.si=6}{b}
\fmfv{l.a=-90,label=$F^\sigma$}{v1}
\fmfv{l.a=-90,label=$F^\alpha$}{v2}
\fmfv{l.a=-90,label=$F^\beta$}{v3}
\fmfv{l.a=-90,label=$F^\delta$}{v4}
\fmf{gluon,right=.7,l.d=10,label=$\hspace{-1cm}\leftarrow p-k_1-k$}{v3,v1}
\fmf{gluon,right=.7,rubout,l.d=10,label=$\leftarrow k_1$}{v4,v2}
\fmf{photon,tension=0}{a,b4}
\end{fmfgraph*}}
\end{align*}
\end{fmffile} 
\caption{\fign{DX} Illustration of \eq{X-gaug-Sig} for the gauged self-energy $\Sigma^\mu$  in the model defined by \eq{Sigma-2gam-1}. Note that in the first graph $k'= k+q=k+p'-p$.}
\end{figure}

\noindent
Equation (\ref{hello}) has the form of a BS equation for $S^\mu$ with the inhomogeneous term $F^\mu$. Evidently,  gauging  the self-energy of \eq{Sigma-2gam-1} generated not only the BS kernel $K_X$, but also an extra inhomogeneous term $\int d^4k/(2\pi)^4 K_X^\mu(p',p;k,k)S(k)$. As a result we get,
\be
S^\mu(p) =\int \frac{d^4k}{(2\pi)^4} G(p',p,k',k) F^\mu(k),  \eqn{G-Gamma}
\ee
where $G$ is the solution of the BS equation $G=G_0+G_0K_X G$ whose pole structure contains full information about  the meson whose 3-photon decay is studied in this paper. Here $F^\mu\ne e\gamma^\mu,$ and therefore \eq{G-Gamma} differs from the form (\ref{Smu-4Greenm}) implied by exact QCD. Moreover, \eq{G-Gamma} is gauge invariant by construction, while \eq{Smu-4Greenm} is not.

\section{Derivation of the \bm{$\Phi\rightarrow 2\gamma$} amplitude.}\label{2gamma}

The amplitude $A^{\mu\nu}$ for the  $\Phi\rightarrow \gamma(q_1,\mu)+\gamma(q_2,\nu)$ transition can be formally derived by gauging  \eq{Amn} for the one-photon decay amplitude $A^\mu$:
\begin{align}
A^{\mu\nu} &=\int d^4x\, e^{-iq_2x}\la0|J^\nu(x)J^\mu(0)|\Phi\ra\nn
&=\left[\la0|J^\mu(0)|\Phi\ra\right]^\nu\nn
&=\int \frac{d^4p}{(2\pi)^4} \mbox{Tr}\left[\bar F^\mu(p)\Phi(P,p)\right]^\nu  \nn
&=\int  \frac{d^4p}{(2\pi)^4} \mbox{Tr}\left[\bar F^{\mu\nu}(p)\Phi(P,p)+\bar F^\mu(p)\Phi^\nu(P,p)\right] 
\eqn{ph-mu-nuA}
\end{align}
which in shorthand is simply
\be
A^{\mu\nu}= \left(\bar F^\mu\Phi\right)^\nu =\bar F^{\mu\nu}\Phi+\bar F^\mu\Phi^\nu .  \eqn{Amn1}
\ee
The term  $\bar F^{\mu\nu}$ is derived by gauging \eq{bFmu}:
\be
\bar F^{\mu\nu}=\left(e\gamma^\mu+S K^\mu\right)^\nu=S^\nu K^\mu+S K^{\mu\nu} . \eqn{Fmunu}
\ee 
Although we do not display momentum variables or integrals over them, these are easily restored at any stage; for example, the full result symbolised by \eq{Fmunu} is 
\begin{align}
[\bar F^{\mu\nu}]^a_b(p_1)&=\int  \frac{d^4k}{(2\pi)^4} \sum_{c,d}\biggl\{[S^\nu]^c_d(k)
[K^\mu]^{da} _{cb}(k,k+q_2;p_1,p_2) \nn
&\hspace{2cm}+S^c_d(k)
[K^{\mu\nu}]^{da} _{cb}(k,k;p_1,p_2)\biggr\} .
 \eqn{Gam-mu-nuA}
\end{align}
The gauged wave function $\Phi^\nu$ is derived by gauging \eq{BS}  (in shorthand $\Phi=G_0K\Phi$) - see Section II C of Ref.\ \cite{KB3} for details:
\begin{align}
\Phi^\nu &=\left(G_0^\nu K+G_0 K^\nu\right)\Phi+G_0K\Phi^\nu  \nn
&=(1-G_0K)^{-1}\left(G_0^\nu K+G_0K^\nu\right)\Phi\nn
&=GG_0^{-1}(G_0^\nu K+G_0K^\nu)\Phi
\nn
&=G(\Gamma_0^\nu G_0K+K^\nu)\Phi=G(\Gamma_0^\nu+K^\nu)\Phi
 \eqn{phi-nuA}
 \end{align}
where 
\be
\Gamma_0^\nu=G_0^{-1}G_0^\nu G_0^{-1}=G_0^{-1}(S_1^{\nu}S_2+S_1S_2^{\nu})G_0^{-1}
\eqn{Gamma0nu}
\ee
 is the sum of the dressed constituent currents. 
Note that $S_1^\nu S_2+S_1 S_2^\nu$ is shorthand for the gauged non-interacting $q\bar q$ propagator
\begin{align}
\left[G_0^\nu\right]^{b_2a_1}_{b_1a_2}
(k_1,k_2;p_1,p_2)
&=(2\pi)^4 \delta^4(k_1-p_1) S^{a_1}_{b_1}(p_1)[S^\nu]_{a_2}^{b_2}(k_2)
\nn
&+(2\pi)^4\delta^4(k_2-p_2)
[S^\nu]^{a_1}_{b_1}(p_1)S_{a_2}^{b_2}(p_2). \eqn{G0mu}
\end{align}
In \eq{G0mu}  the gauged single quark propagators $[S^\nu]_{a_2}^{b_2}(k_2)$ and $
[S^\nu]^{a_1}_{b_1}(p_1)$ are displayed as functions of $k_2$ and $p_1$ (and not $p_2$ and $p_1$) because these are the momenta associated with the {\it incoming} quark lines.

Taking into account \eq{SmuFG}, the two-photon decay amplitude, (\ref{Amn1}), can thus be rewritten as
\begin{align}
A^{\mu\nu}&=\bar F^{\mu\nu}\Phi+\bar F^\mu G(\Gamma_0^\nu+K^\nu)\Phi \nn
&=\bar F^{\mu\nu}\Phi+ S^\mu(\Gamma_0^\nu+K^\nu)\Phi \eqn{Amn2}
\end{align}
which with exposed momentum variables reads
\begin{align}
A^{\mu\nu}& =\int  \frac{d^4p}{(2\pi)^4} \bar F^{\mu\nu}(p)\Phi(P,p) \nn
&+\int \frac{d^4k_1\, d^4p_1}{(2\pi)^8}S^\mu(k)\big[\Gamma_0^\nu(k_1,k_1+q_1;p_1,p_2)+K^\nu(k_1,k_1+q_1;p_1,p_2)\big]\Phi(P,p_1). \eqn{Amn3}
\end{align}
One can check that momentum conservation, $q_2-k_1+k_2+p_1-p_2=0$ (with $k_2=k_1+q_1$), is satisfied in $\Gamma_0^\nu(k_1,k_2,p_1,p_2)$ and $K^\nu(k_1,k_2,p_1,p_2)$  involved in the \eq{Amn3}. Note that the incoming $q_2$ is involved in the momentum conservation relation because the operators $\Gamma_0^\nu$ and $K^\nu$ are meant to be gauged using the current operator $J^\nu$ which carries an incoming momentum of $q_2$.

Using \eq{Gamma0nu} and the fact that $S^\mu = S \Gamma^\mu S = \Gamma^\mu G_0$, one can write
\be
S^\mu\Gamma_0^\nu = \Gamma^\mu S \Gamma^\nu + \Gamma^\nu S \Gamma^\mu \eqn{SG0}
\ee
so  that
\be
S^\mu\Gamma_0^\nu\Phi=
\mbox{Tr}\, (\Gamma^\mu S  \Gamma^\nu 
+\Gamma^\nu S \Gamma^\mu)\Phi . \eqn{tr}
\ee
Substituting this and \eq{Fmunu} into \eq{Amn2}, we obtain the final result for the two-photon decay amplitude:
\begin{align}
A^{\mu\nu}&=\mbox{Tr}\, \left(\Gamma^\mu S  \Gamma^\nu +\Gamma^\nu S \Gamma^\mu\right) \Phi
+ \left(S^\mu K^\nu+S^\nu K^\mu+S K^{\mu\nu}\right)\Phi. \eqn{2gam-symbola}
\end{align}
This result is written out in full in \eq{2gam-full} and illustrated in \fig{fig:Amn}.
The first terms on the RHS of \eq{2gam-symbola}
may appear like the case of no exchange currents, $K^\mu=K^\nu=K^{\mu\nu}=0$, but this is not the case as $S^\mu$ and  $S^\nu$ in these lines contain $K^\mu$ and $K^\nu$. 

In the same way, by gauging the expression (\ref{SmuFG}) for $S^\mu$ (i.e. attaching the current $J^\nu$),
one obtains the four-point Green function for quark-antiquark annihilation into two photons,
\be
[S^{\mu\nu}(l)]^{a_1}_{a_2}=\left\{\int  \frac{d^4p}{(2\pi)^4}[\bar F^\mu  (p)]_{b_2}^{b_1}G^{b_2a_1}_{b_1a_2}(P,p,l)\right\}^\nu.
\ee 
This is done explicitly in \eq{Smunu-expl}. The result has the  form of \eq{2gam-full} but with the bound state wave function $\Phi$  replaced by the  two-body quark-antiquark Green function $G$:
\begin{align}\label{Smunu}
&[S^{\mu\nu}(l)]^{a_1}_{a_2}=
\int d^4x\,d^4y\,e^{-i(q_1x+q_2y+lz)}\la0|J^\mu(x)J^\nu(y)q_{a_2}(0)\bar q^{a_1}(z)|0\ra
\nn
&=\int  \frac{d^4p_1}{(2\pi)^4} [\Gamma^\mu(p_1+q_2)S(p_1+q_2)\Gamma^\nu(p_1)]_{b_2}^{b_1}G^{b_2a_1}_{b_1a_2}(P,p_1,l)
\nn
&+\int  \frac{d^4p_1}{(2\pi)^4} [\Gamma^\nu(p_1+q_1)S(p_1+q_1)\Gamma^\mu(p_1)]_{b_2}^{b_1}G^{b_2a_1}_{b_1a_2}(P,p_1,l)
\nn
&+\int  \frac{d^4p_1\,d^4k}{(2\pi)^8}  [S^\mu(k) K^\nu(k,k+q_1;p_1,p_2)
+S^\nu(k)K^\mu(k,k+q_2;p_1,p_2)]_{b_2}^{b_1}G^{b_2a_1}_{b_1a_2}(P,p_1,l)
\nn
&+\int  \frac{d^4p_1\,d^4k}{(2\pi)^8}  [S(k)
K^{\mu\nu}(k,k;p_1,p_2) ]_{b_2}^{b_1}G^{b_2a_1}_{b_1a_2}(P,p_1,l) .
\end{align}

\section{Derivation of the \bm{$\Phi\rightarrow 3\gamma$} amplitude.}\label{3gamma}

The amplitude $A^{\mu\nu\sigma}$ for the $\Phi\rightarrow \gamma(q_1,\mu)+\gamma(q_2,\nu)+\gamma(q_3,\sigma)$ transition can be formally derived by gauging \eq{2gam-symbola} for the two-gamma decay amplitude $A^{\mu\nu}$:
\begin{subequations}\label{3gam-symb}
\begin{align}
A^{\mu\nu\sigma} &=\int d^4y\,d^4x\,e^{-i(q_3y+q_2x)}\la0|TJ^\sigma(y)J^\nu(x)J^\mu(0)|\Phi\ra\nn
&=\left[\int d^4x\,e^{-iq_2x} \la0|J^\nu(x)J^\mu(0)|\Phi\ra\right]^\sigma
\nn[1mm]
&=
\bigl[\mbox{Tr}\,\left(\Gamma^\mu S  \Gamma^\nu + \Gamma^\nu S \Gamma^\mu\right) \Phi
+\left(S^\mu K^\nu+S^\nu K^\mu+S K^{\mu\nu}\right)\Phi \bigr]^\sigma
\nn[2mm]
&=
\mbox{Tr}\, \left(\Gamma^\mu S \Gamma^\nu +\Gamma^\nu S \Gamma^\mu\right) \Phi^\sigma
+\left(S^\mu K^\nu+S^\nu K^\mu+S K^{\mu\nu}\right)\Phi^\sigma    \eqn{C1a}
\\
&\hspace{5mm} +
\mbox{Tr}\, \bigl(\Gamma^\mu S  \Gamma^\nu 
+ \Gamma^\nu S \Gamma^\mu \bigr)^\sigma\Phi  \eqn{C1b} \\
&\hspace{5mm}+\bigl(S^\mu K^\nu+S^\nu K^\mu+S K^{\mu\nu}\bigr)^\sigma\Phi .  \eqn{C1c}
\end{align}
\end{subequations}
Line  (\ref{C1a}) can be simplified by using  \eq{phi-nuA} for the gauged bound state wave function $\Phi^\sigma$, and the expression (\ref{Smunu}) for $S^{\mu\nu}$, which in symbolic form reads
\begin{align}\label{Smunu-symb}
S^{\mu\nu} &=\bigl( \Gamma^\mu S  \Gamma^\nu + \Gamma^\nu S \Gamma^\mu 
+S^\mu K^\nu+S^\nu K^\mu +S K^{\mu\nu} \bigr) G .
\end{align}
One obtains
\begin{align}
\mbox{Tr}\, & \left(\Gamma^\mu S \Gamma^\nu +\Gamma^\nu S \Gamma^\mu\right) \Phi^\sigma
+
\left(S^\mu K^\nu+S^\nu K^\mu+S K^{\mu\nu}\right)\Phi^\sigma\nn
&=S^{\mu\nu}\left(\Gamma_0^\sigma +K^\sigma\right)\Phi .
\end{align}
Thus the sum of lines  (\ref{C1a}) and  (\ref{C1c}) is
\begin{align}
S^{\mu\nu}&\left(\Gamma_0^\sigma + K^\sigma\right)\Phi
+\left(S^\mu K^\nu+S^\nu K^\mu+S K^{\mu\nu}\right)^\sigma\Phi
\nn[2mm]
&=S^{\mu\nu}\Gamma_0^\sigma \Phi + S K^{\mu\nu\sigma }\Phi+
\sum_{cycle}(S^\mu K^{\nu\sigma}+S^{\mu\nu}K^\sigma)\Phi
\end{align}
where the sum is over cyclic permutations of $(\mu\nu\sigma)$:
\begin{align}
\sum_{cycle}(S^\mu K^{\nu\sigma}+S^{\mu\nu}K^\sigma)
&=S^\mu K^{\nu\sigma} + S^\nu K^{\sigma\mu} + S^\sigma K^{\mu\nu}\nn
&\hspace{5mm}+S^{\mu\nu}K^\sigma + S^{\nu\sigma}K^\mu + S^{\sigma\mu}K^\nu.
\end{align}
Note that
\begin{align}
S^{\mu\nu} &=\left( S \Gamma^\mu S\right)^\nu \nn
&= S\left(\Gamma^\mu S \Gamma^\nu + \Gamma^\nu S \Gamma^\mu + \Gamma^{\mu\nu}\right) S. \eqn{Smunu2}
\end{align}
Analogously to  \eq{SG0}, one can use \eqs{Gamma0nu} and (\ref{Smunu2}) to write
\begin{align}
S^{\mu\nu}\Gamma_0^\sigma \Phi &=\mbox{Tr}\, \big[ \left(\Gamma^\mu S \Gamma^\nu + \Gamma^\nu S \Gamma^\mu + \Gamma^{\mu\nu}\right) S\Gamma^\sigma\nn
&\hspace{4mm} + \Gamma^\sigma S \left(\Gamma^\mu S \Gamma^\nu + \Gamma^\nu S \Gamma^\mu + \Gamma^{\mu\nu}\right) \bigr]\Phi,
\end{align}
while line (\ref{C1b}) evaluates to
\begin{align}
\mbox{Tr}\left(\Gamma^\mu  S \Gamma^\nu 
+ \Gamma^\nu S \Gamma^\mu\right)^\sigma\Phi 
&=\mbox{Tr} \bigl(\Gamma^{\mu\sigma} S\Gamma^\nu + \Gamma^\mu S \Gamma^\sigma S \Gamma^\nu
+ \Gamma^\mu S \Gamma^{\nu\sigma}\nn
&\hspace{6mm}+\Gamma^{\nu\sigma} S\Gamma^\mu + \Gamma^\nu S \Gamma^\sigma S \Gamma^\mu
+ \Gamma^\nu S \Gamma^{\mu\sigma} \bigr)\Phi.
\end{align}
Adding the last two equations gives
\begin{align}
S^{\mu\nu}\Gamma_0^\sigma \Phi+\mbox{Tr}\left(\Gamma^\mu  S \Gamma^\nu 
+ \Gamma^\nu S \Gamma^\mu\right)^\sigma\Phi 
&=\sum_{perm}\mbox{Tr}\,  \Gamma^{\mu} S\Gamma^\nu S \Gamma^\sigma \Phi\nn
&+\sum_{cycle}\mbox{Tr}\left( \Gamma^{\mu\nu} S\Gamma^\sigma + \Gamma^\sigma S \Gamma^{\mu\nu}\right)\Phi
\end{align}
where the first sum is over all permutations of the photon polarisation indices $(\mu,\nu,\sigma)$, while the second is only over the cyclic permutations.
Combining these results gives
\begin{align}
A^{\mu\nu\sigma} &=\sum_{perm}\mbox{Tr}\,  \Gamma^{\mu} S\Gamma^\nu S \Gamma^\sigma \Phi
+\sum_{cycle}\mbox{Tr}\left( \Gamma^{\mu\nu} S\Gamma^\sigma + \Gamma^\sigma S \Gamma^{\mu\nu}\right)\Phi\nn
&+ 
\sum_{cycle}(S^{\mu\nu}K^\sigma + S^\mu K^{\nu\sigma})\Phi + S K^{\mu\nu\sigma }\Phi .    \eqn{Amns-Smunu}
\end{align}
Finally, using \eq{Smunu2} in the above gives
\begin{align}
A^{\mu\nu\sigma} &=\sum_{perm}\mbox{Tr}\,  \Gamma^{\mu} S\Gamma^\nu S \Gamma^\sigma \Phi
+\sum_{cycle}\mbox{Tr}\left( \Gamma^{\mu\nu} S\Gamma^\sigma + \Gamma^\mu S \Gamma^{\nu\sigma}\right)\Phi\nn
&+ 
\sum_{perm}   S \Gamma^\mu S \Gamma^\nu S K^\sigma \Phi
+\sum_{cycle} \bigl( S \Gamma^{\mu\nu} S K^\sigma + S\Gamma^\mu S K^{\nu\sigma} \bigr)\Phi + S K^{\mu\nu\sigma }\Phi .
\end{align}
This result is illustrated in \fig{fig:Amns}.

As emphasised throughout this paper, the gauge invariance of our derived photon decay amplitude  relies on the same BS kernel $K$ being used to calculate all the components making up the amplitudes. In the above expression for $A^{\mu\nu\sigma}$, the only such component whose $K$ dependence has not previously been discussed is the two-photon vertex $\Gamma^{\mu\nu}$. To reveal this dependence, we equate two expressions for $S^{\mu\nu}$; one obtained by gauging \eq{SmuFG}:

\begin{align}
S^{\mu\nu} &=\left( \bar F^\mu G\right)^\nu =(S K^\mu)^\nu G+\bar F^\mu G^\nu\nn
&=(S^\nu K^\mu+S K^{\mu\nu}) G+\bar F^\mu G \left(\Gamma_0^\nu+K^\nu\right) G \nn
&=(S^\nu K^\mu+S K^{\mu\nu}) G+ S^\mu \left(\Gamma_0^\nu+K^\nu\right) G \nn
&=\left(S^\nu K^\mu+S K^{\mu\nu} + \Gamma^\mu S\Gamma^\nu + \Gamma^\nu S\Gamma^\mu+S^\mu K^\nu\right) G  \eqn{Smunu-expl}
\end{align}
[which is just \eq{Smunu} in symbolic form], and the other by gauging \eq{Gammamudef}: 
\be
S^{\mu\nu} =\left(\Gamma^\mu S\Gamma^\nu + \Gamma^\nu S\Gamma^\mu + \Gamma^{\mu\nu}\right) G_0 .
\ee
As $G G^{-1}_0=1+G_0 T$ [see \eq{BSG}], one obtains
\begin{align}
\Gamma^{\mu\nu} &=\left( \Gamma^\mu S\Gamma^\nu + \Gamma^\nu S\Gamma^\mu\right) G_0 T +
\left(S^\mu K^\nu +S^\nu K^\mu+S K^{\mu\nu}\right) (1+G_0 T).
\end{align}
As discussed in the main text, a feature of this result is the appearance of the $q\q$ $t$ matrix $T$ whose $K$ dependence is given by the BS equation for $T$, \eq{BST}.

\section{Derivation of the WT identity for \bm{$S^{\mu\nu}$}}\label{WTI}
Here we show that the expression derived  for the EM four-point Green function $S^{\mu\nu}$ in \eq{Smunu-expl}, and given in full notation in \eq{Smunu}, satisfies the WT identity that relates its longitudinal part $q_2{}_\nu S^{\mu\nu}$ to the dressed quark EM three-point function $S^\mu$. To make the derivation more clear, we use the symbolic form of \eq{Smunu}:
\begin{align}\label{Smunu-symb-1}
S^{\mu\nu} &= \bigl( \Gamma^\mu S \Gamma^\nu + \Gamma^\nu S \Gamma^\mu
+S^\mu K^\nu+S^\nu K^\mu+S K^{\mu\nu}\bigr) G \nn[1mm]
&=\bigl(S^\mu \Gamma_0^\nu+S^\mu K^\nu+S^\nu K^\mu+S K^{\mu\nu} \bigr) G,
\end{align}
where we have used \eq{SG0}. As all the gauged operators on the RHS of \eq{Smunu-symb-1} are constructed with the gauging of equations method, they themselves are gauge invariant and satisfy the following WT identities:
\begin{align}
q_2{}_\nu\Gamma_0^\nu=-[\hat e,G_0^{-1}],\hspace{8mm}
q_2{}_\nu K^\nu=[\hat e,K],\hspace{8mm}
q_2{}_\nu K^{\mu\nu}=[\hat e,K^\mu]  \eqn{hat-e-com}
\end{align}
where $\hat e$ is defined in \eq{hat-e}.
In our notation the WT identity for the EM three-point function $S^\nu$ has the symbolic form
\be
q_2{}_\nu S^\nu =-S\hat e     \eqn{Sn-S}
\ee
since
\be
q_2{}_\nu S^\nu(p_1) =ie[S(p_1)-S(p_1+q_2)]=-\int \frac{d^4k_1}{(2\pi)^4} S(k_1)\hat e(k_1,k_1;p_1,p_2) .  
\ee
Then using the WT identities of  \eqs{hat-e-com} and (\ref{Sn-S}), one obtains the WT identity for $S^{\mu\nu}$:
\be
q_2{}_\nu S^{\mu\nu}=-S^\mu\hat e  .\eqn{wti-forSmn}
\ee
Indeed,
\begin{align}\label{}
q_2{}_\nu S^{\mu\nu}
&= q_2{}_\nu \bigl[  S^\mu\left( \Gamma_0^\nu+K^\nu\right)+S^\nu K^\mu +S K^{\mu\nu} \bigr]G
\nn
&=\bigl\{-S^\mu[\hat e,G^{-1}] -S\hat e K^\mu+S [\hat e,K^\mu]\bigr\}G
\nn
&=\bigl(-S^\mu\hat e G^{-1}+S^\mu G^{-1}\hat e -S K^\mu \hat e\bigr)G
\nn
&=\bigl[-S^\mu\hat e G^{-1}+\bigl(\bar F^\mu-S K^\mu\bigr) \hat e \bigr] G
\nn
&=\bigl( -S^\mu\hat e G^{-1} +\gamma^\mu \hat e\bigr) G=-S^\mu\hat e,
\end{align}
where we have used  \eqs{SmuFG}, (\ref{bFmu}) and (\ref{Gammamu-2b}),
and the fact that
\be
\gamma^\mu \hat e=0.
\ee

It is essentially a universal theorem that a WT identity for a Green function implies current conservation for a corresponding on-shell amplitude. Thus the WT identity of \eq{wti-forSmn} for $S^{\mu\nu}$ implies current conservation for the corresponding two-photon decay amplitude $A^{\mu\nu}$:
\be
q_2{}_\nu A^{\mu\nu}=0  .\eqn{cc-forAmn}
\ee
One can  prove this explicitly as follows. It is clear from the above considerations that
\be
A^{\mu\nu} = S^{\mu\nu} G^{-1} \Phi. 
\ee
Choosing the centre of mass system for convenience, we note that $G$ has a bound state pole at the total energy $P^0=M_\Phi$, so that $G^{-1}\Phi=0$; however, $S^{\mu\nu} G^{-1} \Phi\ne0$ because
$S^{\mu\nu}$ itself has a pole at $P^0=-q_1^0-q_2^0=M_\Phi$ which cancels the zero of $G^{-1}$. 
The WT identity of \eq{wti-forSmn} then gives
\be
q_2{}_\nu A^{\mu\nu} = -S^{\mu}\hat e G^{-1} \Phi
\ee
where the $S^\mu$ has the bound state pole at $-q_1^0=M_\Phi$ which therefore cannot cancel the zero of $G^{-1}$ at $q_1^0-q_2^0=M_\Phi$. Hence $q_2{}_\nu A^{\mu\nu}=0$.

\input{3g_2.bbl}

\end{document}

%% file: 3g_2.bbl
%